\documentclass[reprint,aps]{revtex4-1}
\usepackage{amssymb,amscd,amsmath,amsthm}
\usepackage{latexsym,amstext}
\usepackage{dcolumn}
\usepackage{color}

\def\be{\begin{equation}}
\def\ee{\end{equation}}
\def\bea{\begin{eqnarray}}
\def\eea{\end{eqnarray}}
\def\bean{\begin{eqnarray*}}
\def\eean{\end{eqnarray*}}
\def\ds{\displaystyle}
\def\be{\begin{equation}}
\def\ee{\end{equation}}

\def\N{{\mathbb N}}
\def\Z{{\mathbb Z}}

\def\O{\Omega}

\begin{document}

\title{Groups, Jacobi functions and rigged Hilbert spaces}

\author{E. Celeghini$\,^{\&,}$}
 \email{celeghini@fi.infn.it}
 \affiliation{Dipartimento di Fisica, Universit\`a di Firenze \\and  INFN-Sezione di Firenze\\
150019
Sesto Fiorentino, Firenze, Italy.}

\author{M. Gadella}
 \email{manuelgadella1@gmail.com}

 \author{M. A. del Olmo}
 \email{marianoantonio.olmo@uva.es}

\affiliation{$^\&$Departamento de F\'{\i}sica Te\'orica, At\'omica y \'Optica, \\
 IMUVA -- Mathematical Research Institute,\\ 
 Universidad de Valladolid, E-47005 Valladolid, Spain.
}

\date{\today}
\begin{abstract}

This paper is a contribution to the study of the relations between special functions, Lie algebras and rigged Hilbert spaces. The discrete indices and continuous variables of special functions are in correspondence with the representations of their algebra of symmetry, that  induce discrete and continuous bases  coexisting on  a rigged Hilbert space supporting the representation. Meaningful operators are shown to be continuous on the spaces of test vectors and its dual. Here, the chosen special functions,  called ``Algebraic Jacobi Functions''  are
related to the Jacobi polynomials  and the Lie algebra is $su(2,2)$.
These  functions with $m$ and $q$ fixed, also 
exhibit a  $su(1,1)$-symmetry. Different  discrete and continuous bases are introduced. An extension in the spirit of the associated Legendre polynomials and the spherical harmonics is presented introducing the  ``Jacobi Harmonics'' 
that are a generalization of the spherical harmonics to the three-dimensional  hypersphere $S^3$.

\begin{description}
\item[PACS numbers]
02.20.Sv, 02.30.Gp,  02.30.Px, 02.30.Tb, 03.65.-w, 03.65.Ca
\end{description}

\noindent
{\bf Keywords:} Special functions, Jacobi polynomials, Lie groups and algebras, rigged Hilbert spaces, harmonic analysis
\end{abstract}

%%%%%%%%%%%%%%%%%%%%%%%%%%%%%%%

\maketitle

%%%%%%%%%%%%%%%%%%%%%%%%%%%%%%%%%%%%%%%%%%%%%%%%%%%%%%%%%%%
%%%%%%%%%%%%%%%%%%%%%%%%%%%%%%%%%%%%%%%%%%%%%%%%%%%%%%%%%%%

\section{Introduction}

This article belongs to a series of papers intended to give a unified picture of the connections between Lie algebras, special functions and rigged Hilbert spaces. The latter share with the related Lie algebras a common property, which is the need to deal simultaneously with operators, with discrete and continuous spectrum and with their  eigenvectors while special functions depend on discrete indices and on continuous variables.
All of them have thus a   common general frame.

One of the main interests of this line of research is to give a consistent description of Quantum Mechanics, that demand the coexistence of operators with discrete spectrum (like the Hamiltonian for the harmonic oscillator)   and continuous spectrum (like the position and momentum operators, etc.), along to their eigenvalues. 
%These eigenvalues are relevant for various purposes, one of the most important being 
%the spectral decomposition of these operators.
% following the Dirac formulation.

In previous works, we have obtained the  algebraic structure associated to some orthogonal polynomials such as Hermite, Legendre, Laguerre, associated Laguerre polynomials or spherical harmonics \cite{CO,CO1,CO2,CO3}. Each class of orthogonal polynomials is associated to a particular representation of a specific  Lie group. This is the case of Hermite polynomials with Heisenberg-Weyl group or associated Legendre polynomials and spherical harmonics with $SO(3,2)$.   Spaces supporting the representation of these associated groups include discrete as well as continuous basis as eigenvectors of the Lie algebra generators or some functions constructed with these generators.  Hilbert spaces are not sufficient to bear a representation of these groups and their corresponding algebras, as they do not have continuous bases. These bases are quite often used by physicists, so that it would be convenient to have representation spaces including both discrete and continuous basis. For this reason, we move to rigged Hilbert spaces where different cardinalities find a formal arrangement.

A rigged Hilbert space (RHS) is a triplet of spaces
\begin{equation*}\label{1}
\Psi\subset \mathcal H\subset \Psi^\times\,,
\end{equation*}
where, $\mathcal H$ is an infinite dimensional separable Hilbert space. The space  $\Psi$ is a dense subspace endowed with a locally convex topology (in many cases produced by an infinite set of norms that include the Hilbert space norm) stronger than the topology induced by the Hilbert space topology (the canonical injection is continuous). It is sometimes called the {\it space of test vectors} or even the {\it  test space}. The space $\Psi^\times$ is the (anti)-dual space of $\Psi$, i.e., the space of all (anti)-linear continuous functionals on $\Psi$, endowed with a topology compatible with the dual pair structure (quite often the weak topology). The action of $F\in\Psi^\times$ on each vector $\varphi\in\Psi$ gives a complex number %that we 
often denoted as $\langle \varphi|F\rangle$, %in order to keep up with 
following the Dirac notation.

It is not our intention here to review properties and applications of RHS, the reader may visit for this purpose the extensive literature on the subject \cite{B,R,ANT,MEL,GG,GG1,GG2,BG,CG}, but we would like to mention that 
recently they have received renewed attention, see \cite{fe09,1,2,3,4,HJ3}, including their relations with partial inner products \cite{5} and frames \cite{6}.   We limit us  to remember that our primary purpose is to provide of mathematical sense to the Dirac formulation of quantum mechanics and its continuous basis of eigenvectors of observables. Also, it is important to recall that if $\mathcal O$ is an operator on $\mathcal H$, such that $\Psi$ reduces its adjoint operator 
$\mathcal O^\dagger$ (i.e., $\mathcal O^\dagger\Psi\subset\Psi$) and  $\mathcal O^\dagger$ is continuous with the topology on $\Psi$, then, it is possible to extend $\mathcal O$ into $\Psi^\times$ by means of the {\it duality formula}\:
\begin{equation}\label{2}
\langle \mathcal O^\dagger \varphi |F\rangle =\langle\varphi |\hat {\mathcal O} F\rangle\,,\qquad \forall\,\varphi\in\Psi\,,\;\;\forall\,F\in\Psi^\times\,.
\end{equation}
This expression is also valid if $\mathcal O$ is self-adjoint.
Hereafter for the sake of the simplicity we will also denote by the same symbol the operator $(\mathcal O)$  and its extension 
$(\hat {\mathcal O})$.

Two RHS  $\Psi\subset\mathcal H\subset \Psi^\times$ and $\Phi\subset\mathcal L\subset\Phi^\times$ are unitarily equivalent if there is a unitary mapping $U:\mathcal H\longmapsto\mathcal L$ such that $U\Psi=\Phi$ is a one-to-one onto mapping preserving topologies, i.e., $U$ and its inverse, $U^{-1}$, are continuous. As a consequence, the duality formula \eqref{2}
\[
\langle U\varphi|UF\rangle=\langle\varphi|F\rangle\,,\qquad  \forall \varphi\in\Psi\,,\;\; \forall F\in\Psi^\times\,,
\]
 extends $U$ to a continuous one-to-one mapping from $\Psi^\times$ onto $\Phi^\times$ with  continuous inverse, i.e. we have the diagram
 \[
\begin{array}{lllclll}
&\Psi&\subset &{\mathcal H}&\subset&\Psi^\times
\\[0.2cm]
  & \hskip-0.35cm  {U}\downarrow && \hskip-0.30cm {U}\downarrow &&\hskip-0.30cm {U}\downarrow
\\[0.2cm]
&\Phi &\subset  & \mathcal L &\subset   &\Phi^\times
\end{array}\,.
\]
A typical situation comes when the former is an abstract RHS and the other is a space in which $\Phi$ and $\mathcal L$ are spaces of functions. We say that the second is a representation of the former.

In previous articles, we have analysed  the connections between  Hermite and Laguerre functions with the fractional Fourier transformation and signal theory \cite{CGO}, discrete and continuous basis for the representations of $SO(3,2)$ and the role of spherical harmonics \cite{CGO1}, and similar discussions concerning other groups and their algebras such as $SO(2)$ \cite{CGO2}, some considerations on harmonic analysis on the circle \cite{ENT}, or $SU(2)$, associated Laguerre polynomials \cite{CGO3} and Zernike polynomials, used in optics, related to the group $SU(1,1)\otimes SU(1,1)$ \cite{CGO19}. In Ref.~\cite{CGO19ax} the interested reader can find a review devoted to   the relation between  Lie groups, special functions and rigged Hilbert spaces.

From the formal point of view, operators with purely discrete spectrum are connected to discrete indices of special functions. They are continuous on certain test space of fuctions $\Phi$ of a given RHS and determine the unitary irreducible representations (UIR) of Lie algebras. Analogously, operators with continuous spectrum are related with the dependence of special functions with respect to continuous real variables. They are continuous on $\Phi$ and, therefore, may be extended by continuity to the anti-dual space $\Phi^\times$, where their eigenvectors are well defined. Recall that eigenvectors for self-adjoint operators corresponding to eigenvalues in the continuous spectrum are not normalizable.

In this paper, we propose a detailed  study of  the ``Algebraic Jacobi Functions'' related to the Jacobi polynomials, that we
   introduced in \cite{CO2,COV}.  The formal structure underlying Jacobi polynomials is discussed. In particular,  we analyse topological properties of the  relevant ladder operators, including the generators of the Lie algebra.  The motivation of our work is twofold. Firstly, as we mention before, we have studied other orthogonal polynomials, like Hermite, Laguerre, Legendre, Associate Legendre, Spherical Harmonics and all of them are related to the Jacobi polynomials \cite{CO,CO1}. On the other hand, the Jacobi polynomials are involved in  many applications in different science areas. For instance,  in physics in the study of quantum superintegrable systems as solutions of such as systems \cite{CNO,TTW}, in the study of the rotations on the the space through the $d_j$-Wigner rotation matrices \cite{wigner,biedenharn},  in signal processing like in the study of solutions of transverse vibrations of nonuniform homogeneous beams and plates \cite{caruntu}, or  in the data compression of electrocardiograms  \cite{TWLH} in the medicine field or properly in mathematics as approximated  solutions of singular integral equations \cite{kps}, in numerical solutions  of integro-differential-difference equations \cite{bbcs}, and in the obtention of Jacobi approximations in Hilbert spaces to  solve singular differential equations \cite{guo00}, or in spectral methods for solving partial differential equations  on unbounded domains \cite{sw09}.

The paper is organized as follows. Next  Section presents a brief revision of  known material: the Algebraic Jacobi Functions together with some of their relevant properties and their symmetries, the Lie algebra $su(2,2)$. In Section \ref{basicspaces} we introduce an infinity family of rigged Hilbert spaces $\{\Phi_{m,q}\subset \mathcal L_{m,q} \subset (\Phi_{m,q})^\times\}$  labelled by two (fixed) integer or half-integer indices $(m,q)$ where $ \mathcal L_{m,q}=L^2[-1,1]$  and we study the operators acting inside 
$\Phi_{m,q}$ and into different spaces $\Phi_{m,q}$ all of them belonging to  $su(2,2)$ or to its Universal Enveloping Algebra UEA$[su(2,2)]$. It is proved that these operators are continuous 
and can be extended continuously to the dual spaces.    
The discrete and continuous bases are studied in Section \ref{discretecontinuousbases}. For that purpose an abstract  Hilbert space $\mathcal H_{m,q}$ is introduced via a unitary map, $U$, 
from $\mathcal H_{m,q}$  into $\mathcal L_{m,q}$. In this way we obtain a new family of abstract rigged Hilbert spaces ($\Psi_{m,q}\subset \mathcal H_{m,q} \subset (\Psi_{m,q})^\times$)  unitarily equivalent to the former one, In this new family we introduce a discrete basis ($\{|j,(m,q)\rangle\}_{j\ge \sup(|m|,|q|)}^\infty$), while a discrete-continuous basis ($\{|x,(m,q)\rangle\}_{x\in [-1,1]}$) is associated with each element of the original family of RHS. Also the $su(2,2)$ operators are translated to the new RHS  as well as their properties by means of $U$. Section \ref{totalspaces} is devoted to consider all together the AJF in a ``total space'' with a inner product, for that we extend the AJF following the procedure used with the Associated Legendre polynomials to obtain the Spherical Harmonics, but here we have to consider two angular variables, $\theta$ and $\xi$ associated the discrete labels $m$ and $q$,  respectively, and not only one like in the Spherical Harmonics.  We also  prove that the $su(2,2)$ operators %, amended accordingly, 
 are continuous and also continuously extended to the dual. Moreover we obtain two RHS one associated to the integer values of $(j,m,q)$ and the other one for  the half-integer values. After a section of conclusions we have added three appendices.
%%%%%%%%%%%%%%%%%%%%%%%%%%%%%%%%%%%%%%%%%%%%%%%%%%%%%%%%%%%
%%%%%%%%%%%%%%%%%%%%%%%%%%%%%%%%%%%%%%%%%%%%%%%%%%%%%%%%%%%

\section{Algebraic Jacobi Functions: an overview}\label{overwiew}

The Jacobi polynomials of order $n\in\mathbb N$, $J^{\alpha,\beta}_n(x)$, are usually defined as \cite{szego,askey}
\begin{widetext}
%\begin{equation}\label{3}
%J^{(\alpha,\beta)}_n(x) := \ds \sum_{s=0}^n \left(\begin{array}{c} n+\alpha \\ s \end{array}\right) \left(\begin{array}{c} n+\beta \\ n-s \end{array}\right) \left(\frac{x+1}{2} \right)^s \left(\frac{x-1}{2} \right)^{n-s}\,,
%\end{equation}
\begin{equation}\label{3}
J^{(\alpha,\beta)}_n(x) := \ds \sum_{s=0}^n \frac{\Gamma ( n+\alpha +1)}{ \Gamma (s+1) \Gamma (n+\alpha -s+1) } \frac{\Gamma ( n+\beta +1)}{ \Gamma (n-s+1) \Gamma (\beta +s+1) } \left(\frac{x+1}{2} \right)^s \left(\frac{x-1}{2} \right)^{n-s}\,.
\end{equation}
%where we have used the generalised binomial coefficient
%\begin{equation*}\label{04}
%\left(\begin{array}{c} a \\ s \end{array} \right):=\frac{\Gamma(a+1)}{\Gamma(s+1)\,\Gamma(a-s+1)}= 
%\frac{a\cdot(a-1)\cdot(a-2)\cdots (a-(s-1))}{s!}\,,
%\end{equation*}
%where $a$ is an arbitrary number and $s$ a positive integer.
They verify the following second order differential equation
\begin{equation*}\label{03}
\left[ (1-x^2) \frac{d^2}{dx^2} -((\alpha+\beta+2)x+(\alpha-\beta)) \frac{d}{dx} + n(n+\alpha+\beta+1)\right]\,
J_n^{(\alpha,\beta)}(x)=0 \,.
\end{equation*}

%%%%%%%%%%%%%%%%%%%%%%%%%%%%%%%
%%%%%%%%%%%%%%%%%%%%%%%%%%%%%%%
\subsection{Algebraic Jacobi Functions}\label{jacobifunctions}

We now consider the representation spaces  of the Lie algebras $su(2,2)$
and $su(1,1)$ that have as bases the {\it Algebraic Jacobi Functions} (AJF)
 defined   in terms of the Jacobi polynomials $J^{(\alpha,\beta)}_{n}(x)  $ by \cite{CO2}:
\begin{equation}\label{5}
\mathcal J^{m,q}_j(x):= \sqrt{\frac{\Gamma(j+m+1) \Gamma(j-m+1)}{\Gamma(j+q+1) \Gamma(j-q+1)}} \left(\frac{1-x}{2} \right)^{\frac{m+q}{2}} \left(\frac{1+x}{2} \right)^{\frac{m-q}{2}} J^{(m+q,m-q)}_{j-m}(x)\,,
\end{equation}
where 
\begin{equation}\label{6}
j:= n+\frac{\alpha+\beta}{2}\,, \qquad m:= \frac{\alpha+\beta}{2}\,,\qquad q:= \frac{\alpha-\beta}{2}\,.
\end{equation}
or
\begin{equation*}\label{06}
n=j-m\,,\qquad  \alpha=m+q\,,\qquad \beta=m-q\,,
\end{equation*}

Considerations derived from the theory of group representations force the following restrictions  in the above parameters:
\begin{equation}\label{7}
j \ge |m|\,, \quad j\ge |q|\,, \quad 2j\in\mathbb N\,, \quad  j-m\in\mathbb N\,, \quad  j-q\in\mathbb N\,,
\end{equation}
with the parameters $(j,m,q)$  all together integers or half-integers. Conditions \eqref{7} in terms of the original parameters $(n,\alpha,\beta)$ are
\begin{equation*}\label{8}
n\in\mathbb N\,,\quad \alpha,\beta \in \mathbb Z\,, \quad \alpha,\,\beta \ge -n\,, \quad  
\alpha+\beta \ge-n\,.
\end{equation*}
The values of  $\mathcal J^{m,q}_j(x)$  at  the points $x=\pm 1$  are defined by means of the appropriate   limits. In  Appendix A it is shown that 
 $\mathcal J^{m,q}_j(x)$ are real polynomials or ``quasi-polynomials''  for  $-1\leq x\leq 1$ and that
 they have quite more symmetries than the Jacobi polynomials, i.e. 
\begin{equation*}\begin{array}{l}
\label{jacobiigualdades}
\mathcal J^{m,q}_j(x) =\mathcal J^{q,m}_j(x)= (-1)^{m+q} \mathcal J^{-m,-q}_j(x) =
(-1)^{j-m}\,\mathcal J^{m,-q}_j(-x)\,.
\end{array}
\end{equation*}

The  Algebraic Jacobi Functions ${\cal J}_j^{m,q}(x)$ verify the differential equation 
\begin{equation}\label{jacobiequation}
\left[-(1-x^2)\,\frac{d^2}{dx^2}+2\,x\frac{d}{dx}+ 
\frac{2\;m\;q\;x+m^2+q^2}{1-x^2}- j(j+1)\right] \,{\cal J}_j^{m,q}(x)=0\; ,
\end{equation}
\end{widetext}

In addition, for fixed $m$ and $q$ the AJF satisfy the following relations:
\begin{equation}\label{9}
\int_{-1}^1 \mathcal J_j^{m,q}(x) (j+1/2) \, \mathcal J_{j'}^{m,q}(x)\,dx =\delta_{j,j'}
\end{equation}
and
\begin{equation}\label{10}
\sum_{j\ge \sup(|m|,|q|)}^\infty \mathcal J_j^{m,q}(x) (j+1/2) \, \mathcal J_{j}^{m,q}(y) = \delta(x-y)\,.
\end{equation}
In expressions \eqref{9} and \eqref{10}   
%the admissible values of $j$ (see eq.~\eqref{7}) are, in one side $j=0,1,2,\dots$ and on the other $j=1/2,3/2,\dots$ in function of the  fixed values of 
$j,m,q$ are all together integrer or half-integer and satisfy \eqref{7}. 
This shows that for fixed $m$ and $q$ both integer or half-integer, the set of AJF given by $\{ J_j^{m,q}(x)\}_{j\ge \sup(|m|,|q|)}^\infty$ forms an orthogonal basis of the Hilbert space $L^2[-1,1]$. However, the multiplication for $\sqrt{j+1/2}$ of the AJF $\mathcal J_j^{m,q}(x)$ 
allows to obtain the related orthonormal basis of $L^2[-1,1]$ in the framework of RHS. In the following we shall use AJF
to discuss the symmetry operators and the {\it Normalized  Algebraic Jacobi Functions} (NAJF)
\begin{equation}\label{NAJF}
\mathbb J_j^{m,q}(x):= \sqrt{j+1/2}\,\mathcal J_j^{m,q}(x)\end{equation}
 as orthonormal basis. %  of $L^2[-1,1]$.
 %in the framework of the RHS. 
The relations \eqref{9} and \eqref{10} become%, respectively,
 \begin{widetext}
 \begin{equation}\begin{array}{rll}\label{0010}
\langle \mathbb J_j^{m,q}(x)\, | \,\mathbb J_{j'}^{m,q}(x) \rangle &\equiv & \ds 
\int_{-1}^1 \mathbb J_j^{m,q}(x)) \, \mathbb J_{j'}^{m,q}(x)\,dx =\delta_{j,j'}\,,\\[0.4cm]
\ds\sum_{j\ge \sup(|m|,|q|)}^\infty |\mathbb J_j^{m,q}(x) \rangle\, \langle\mathbb J_{j}^{m,q}(y)|
&\equiv&\ds\sum_{j\ge \sup(|m|,|q|)}^\infty \mathbb J_j^{m,q}(x) \, \mathbb J_{j}^{m,q}(y) = \delta(x-y)\,.
\end{array}\end{equation}

Note that the AJF have a direct relation  with the Legendre functions:
\begin{equation*}
P_l(x)={\mathcal J}_l^{0,0}(x)\,,\qquad 
P_l^m(x) =(-1)^m \sqrt{\frac{(l+m)!}{(l-m)!}}\,\mathcal J_l^{m,0}(x)\,,
\end{equation*}
as can be obtained from  \cite{HAND} and \cite{MAG} (p. 213).

%%%%%%%%%%%%%%%%%%%%%
%%%%%%%%%%%%%%%%%%%%%
\subsection{$su(2,2)$ and $su(1,1)$  symmetry algebras on AJF}

The explicit formulae for the  ladder operators, $A_\pm,B_\pm,C_\pm,D_\pm,E_\pm,F_\pm$  on the  Algebraic Jacobi Functions are given in \cite{CO2,COV}. They are  generators of the $su(2,2)$ Lie algebra and their action on the AJF is given by:
\begin{equation}\begin{array}{lll}\label{11}
A_\pm\, \mathcal J^{m,q}_j(x)&=& \sqrt{(j\mp m)(j\pm m+1)}\,\mathcal J^{m\pm 1,q}_j(x)\,,
\\[0.4cm]
B_\pm \, \mathcal J^{m,q}_j(x)&=& \sqrt{(j\mp q)(j\pm q+1)}\, \mathcal J^{m,q\pm 1}_j(x)\,,
\\[0.4cm]
C_\pm\,{\cal J}_j^{m,q}(x)&=&\sqrt{(j+m+\frac12\pm\frac12)(j+q+\frac12\pm\frac12)}\; {\cal J}_{j\pm 1/2}^{m\pm 1/2,\;q\pm 1/2}(x),\\[0.4cm]
D_\pm\,{\cal J}_j^{m,q}(x)&=&\sqrt{(j+m+\frac12\pm\frac12)(j-q+\frac12\pm\frac12)}\; {\cal J}_{j\pm1/2}^{m\pm1/2,\;q\mp 1/2}(x)\,,\\[0.4cm]
E_\pm\,{\cal J}_j^{m,q}(x)&=&\sqrt{(j-m+\frac12\pm\frac12)\,(j+q+\frac12\pm\frac12)}
\;{\cal J}_{j\pm 1/2}^{m\mp 1/2,\;q\pm 1/2}(x),\\[0.4cm]
F_\pm\,{\cal J}_j^{m,q}(x)&=&\sqrt{(j-m+\frac12\pm\frac12)\,(j-q+\frac12\pm\frac12)}\;
\,{\cal J}_{j\pm 1/2}^{m\mp 1/2,\,q\mp 1/2}(x)\,.
\end{array}\end{equation}
\end{widetext}

The three diagonal operators $J,M$ and $Q$\begin{equation}\label{jmqoperators}
(J,\; M ,\; Q)\, {\cal J}_j^{m,q}(x)\,=\; (j, \;m,\; q)\,{\cal J}_j^{m,q}(x)\, ,
\end{equation}
 belong   to the Cartan subalgebra.

The 15-dimensional (conformal)  simple Lie algebra $su(2,2)$  has many applications in physics (see, for instance, \cite{barut,ORW,calixto} and references therein) and is a non-compact real form of the complex algebra $A_3$  \cite{mack,jacobson}.

Note that  the operators $A_\pm$ ($B_\pm$)   change the  label $m$ ($q$) by $\pm 1$  leaving invariant the remaining two labels. They generate the representation $D_j\otimes D_j$ of the maximal compact subalgebra $su(2)\oplus su(2)$ of $su(2,2)$. However, the operators   $C_\pm,\cdots, F_\pm$ change all the $(j,m,q)$
by a half-integer quantity. Integer and half-integer values of $(j,m,q)$ are indeed related to a non-unitary and non-irreducible unique representation  of $SU(2,2)$. 
%As we saw  before in \eqref{7}  the labels of any ${\cal J}_j^{m,q}$ verify  $j\geq|m|$ and $j\geq|q|$.
Obviously the algebra is consistent with relation \eqref{7}.

Furthermore  composing the action of two pairs of operators  \eqref{11} we can 
construct two operators that change $(j,m,q)$ to $(j\pm1, m ,q)$.
Indeed $F_\pm C_\pm$ (equivalently, $C_\pm F_\pm$,  $D_\pm E_\pm$ or
$E_\pm D_\pm$)  are such  operators. These new operators are second
order differential operators but they can be  reduced by means of the Jacobi equation  (\ref{jacobiequation}) to  first order
differential operators when applied to the ${\cal J}_j^{m,q}$ \cite{CO2}.
To define these new algebraic operators, we have to consider separately   specific values of $m$ and $q$:  $j\geq|m| > |q|$,  $j\geq|q| > |m|$ and    $j\geq |m| = |q|$:

1) For the  states such that  $j\geq|m| > |q|$ we can define 
from \eqref{11} and \eqref{jmqoperators}  two conjugate Hermitian operators as
\begin{equation}\label{Kpm}
K_\pm:=   F_\pm C_\pm  \frac{1}{\sqrt{(J+1/2\pm 1/2)^2-Q^2}} ,
\end{equation}
for which the action on the ${\cal J}_j^{m,q}(x)$ does not depend from the value of $q$:
\begin{equation} 
\label{actionK}
 K_\pm\;  {\cal J}_j^{m,q}(x) = \sqrt{(j+1/2\pm 1/2)^2-m^2}\; 
 {\cal J}_{j\pm1}^{m,q}(x) \,.
\end{equation}
Both operators \eqref{Kpm} together with $K_3:=J+1/2$ close a 
$su(1,1)$ Lie algebra
\begin{equation} 
\label{su11}
[K_+,K_-]=-2 K_3,\qquad [K_3,K_\pm]=\pm K_\pm\,,
\end{equation}
i.e.,  the set $\{{\cal J}_j^{m,q}\}^{m,q\, {\rm fixed}}_{j\geq |m| > |q|}$, with $|m|>|q|$,   is  a basis of the UIR of $SU(1,1)$  with Casimir ${\cal C} = m^2 -1/4$.

2) For the ${\mathcal J}_j^{m,q}(x)$ with $j\geq |q|>|m|$, the procedure is analogous. In fact, we only have  to interchange $M \Longleftrightarrow Q$ in the definition of  $K_\pm$. 

3) The third case shows a difficulty for the  definition of the operator $K_-$ as obtained from  \cite{CO2,COV} when $j=|m|=|q|$, since  $K_-$ 
is not well defined in eq.(\ref{Kpm}). The following limit  allows to avoid this difficulty:
\begin{widetext}
\begin{equation}\label{20}
K_- := \lim_{\epsilon \to 0} 
\left[ 
\left((1-X^2) \,D_X+  X\, J+
\frac{(M+\epsilon)(Q+\epsilon)}{J}
\right)
\frac{J}{\sqrt{J^2-(Q+\epsilon)^2}}\; \right]\,,
\end{equation}\end{widetext}
where $X$ and $D_X$ are  the multiplication and derivative operators  
(i.e., $X\,f(x)=x\,f(x)$, $D_X\,f(x)=d\,f(x)/dx$). In Appendix B we study the continuity of the multiplication operator defined in \cite{CO2,COV}.

This limit removes the problem because the action of $K_-$ on ${\mathcal J}_j^{m,q}(x)$ agrees with  \eqref{actionK} for 
$j > |m|= |q|$, while  for 
$j=|m|=|q|$ vanishes because
 \eqref{20}  is the product of two terms, one goes like $\epsilon$ and the other like $\epsilon^{-1/2}$.  
In conclusion, if $|m|\geq |q|$  the action of the $K_\pm$ is given by eqs.~\eqref{actionK}
with $j=|m|, |m|+1, |m|+2 \dots$  and  Casimir invariant 
${\cal C} = m^2-1/4$, while for  $|m| < |q|$, the result is the same after the exchange of $q$ and  $m$.
Thus, the set  $\{{\mathbb J}_j^{m,q}\}^{m,q\, {\rm fixed}}_{j\geq {\rm sup}( |m|, |q|)}$ supports, because of \eqref{0010},  
a UIR of $SU(1,1)$ for any $m,q$.

All these operators are continuous linear mappings between suitable locally convex spaces admitting continuous extensions to dual spaces. The construction of these spaces as well as the discussion on their properties  is the subject of the next section.

%%%%%%%%%%%%%%%%%%%%%%%%%%%%%%%%%%%%%%%%%%%%%%%%%%%%%
%%%%%%%%%%%%%%%%%%%%%%%%%%%%%%%%%%%%%%%%%%%%%%%%%%%%%
\section{Basic spaces and their operators
%relations
}\label{basicspaces}

%%%%%%%%%%%%%%%%%%%%%
%%%%%%%%%%%%%%%%%%%%%
\subsection{The spaces $\Phi_{m,q}$} 

In the previous section, we have mentioned that for fixed $m$ and $q$ (both integer or half-integer) the set of normalized  Jacobi functions    \eqref{NAJF} $\{\mathbb J_j^{m,q}(x)\}_{j\ge \sup(|m|,|q|)}^\infty$ is an orthonormal basis for $L^2[-1,1]$. 
Let us construct a set of spaces $\Phi_{m,q}$ each depending on the labels $m$ and $q$, as follows: Let us take $f(x)\in L^2[-1,1]$, which 
  admits the following expansion:
\begin{equation}\label{22}
f(x)= \sum_ {j\ge \sup(|m|,|q|)}^\infty f_j^{m,q}  
\:\mathbb J_j^{m,q}(x)\,.
\end{equation}
Then by definition, $f(x)\in\Phi_{m,q}$ if and only if for all $r,s=0,1,2,\dots$, we have that
\begin{widetext}\begin{equation}\label{23}
[p^{m,q}_{r,s}(f)]^2:=\sum_ {j\ge \sup(|m|,|q|)}^\infty |f_j^{m,q}|^2\,(j+|m|+1)^{2r}\,(j+|q|+1)^{2s}<\infty\,.
\end{equation}
\end{widetext}
Observe that the mappings $p^{m,q}_{r,s}(f)$ define a family of norms on $\Phi_{m,q}$ and hence a topology on each of the spaces $\Phi_{m,q}$. With this topology, $\Phi_{m,q}$ is a nuclear Fr\`echet space \cite{PIE}. Each of the spaces $\Phi_{m,q}$ contains all the functions in the corresponding orthonormal basis, so that these spaces are all dense in $L^2[-1,1]$. Since the norm on $\Phi_{m,q}$ given by $r=s=0$ is just the norm on $L^2[-1,1]$, the topology in $\Phi_{m,q}$ is finer than the topology inherited from $L^2[-1,1]$, so that the canonical mapping $\Phi_{m,q}\longmapsto L^2[-1,1]$ is continuous. If we add the anti-dual $\Phi_{m,q}^\times$, we have for each of the values of $m$ and $q$ the RHS
\[
\Phi_{m,q}\subset L^2[-1,1] \subset \Phi_{m,q}^\times\,.
\]

%Before proceeding, some comments are in order. Let $\Psi\subset\mathcal H\subset\Psi^\times$ be a RHS and 
Let $\mathcal O$ be a densely defined {\it linear} operator on $ L^2[-1,1]$ such that $\Phi_{m,q}$ reduces $\mathcal O$. This means that $\mathcal O\,\phi\in\Phi_{m,q}$ for all $\phi\in\Phi_{m,q}$, or equivalently that 
$\mathcal O\,\Phi_{m,q}\subset\Phi_{m,q}$, so that 
$\mathcal O:\Phi_{m,q}\longmapsto \Phi_{m,q}$. The operator $\mathcal O$ is continuous with respect to the locally convex topology induced by the family of norms $\{{p}_{r,s}^{m,q}\}_{r,s\in \N}^{m, q\, \text{fixed}}$, if and only if for each pair $(r,s)$, with $r,s=0,1,2,\dots$, there exists a positive constant $K>0$ and a finite number of pairs $(r_1,s_1)$, $(r_2,s_2)$, ..., $(r_k,s_k)$, which depend on $(r,s)$, such that for all $f\in\Phi_{m,q}$, we have  \cite{RS}
\begin{equation}\label{024}
p_{r,s} (\mathcal O f) \le K\,\{  p_{r_1,s_1}(f) +p_{r_2,s_2}(f) + \dots+ p_{r_k,s_k}(f) \}\,,
\end{equation}
where we have omitted the superindices $m$  and $q$ for simplicity. Along the present section, we shall omit these two indices unless necessary.

A similar formula proves the continuity of linear mappings between different locally convex spaces.

%%%%%%%%%%%%%%%%%%%%%
%%%%%%%%%%%%%%%%%%%%%
\begin{widetext}
\subsection{Operators acting  on $\Phi_{m,q}$}

Our first goal is to prove that the operators $J$, $M$ and $Q$ defined in \eqref{jmqoperators} are reduced by all the spaces $\Phi_{m,q}$ and are continuous as linear mappings on these spaces. Let us prove this property for $J$ first. 
%We shall omit the indices $m$ and $q$ from the sum and from the coefficients, in order to alleviate the notation. 
After \eqref{NAJF} and \eqref{jmqoperators}  it is clear that, for any $f(x)\in\Phi_{m,q}$ \eqref{22},  we should define $Jf$ as
\begin{equation*}\label{025}
J\,f(x)\equiv J \sum_j f_j 
\,\mathbb J^{m,q}_j(x)  := \sum_j f_j \,j\,
\mathbb J^{m,q}_j(x)\,.
\end{equation*}
Since
\begin{equation}\begin{array}{lll}\label{026}
[p_{r,s}(Jf)]^2 &=&\ds \sum_j |f_j|^2\,j^2 \, (j+|m|+1)^{2r} (j+|q|+1)^{2s} \\[0.4cm]
&\le &\ds \sum_j |f_j|^2\,(j+|m|+1)^{2(r+1)} (j+|q|+1)^{2s}<\infty\,,
\end{array}\end{equation}
(for all $r,s=0,1,2,\dots$), we  get  that $(Jf)(x)\in \Phi_{m,q}$ for any $f(x)\in \Phi_{m,q}$. In addition,
\begin{equation}\label{027}
[p_{r,s}(Jf)]^2 \le \sum_{j} (j+|m|+1)^{2r+2}(j+|q|+1)^{2s} |f_{j}|^2 = p^2_{r+1,s}(f)\,,
\end{equation}%\end{widetext}
which proves the continuity of $J$. On $\Phi_{m,q}$ the operators $M$ and $Q$ are trivial, since they just multiply by a constant, either $m$ or $q$, respectively. Then, these operators are trivially continuous on all $\Phi_{m,q}$.

The operators $K_\pm$ \eqref{actionK} are also linear and continuous on the spaces $\Phi_{m,q}$. 
Let us write $(K_+f)(x)$
for any $f(x)\in\Phi_{m,q}$ by   using the expansion of  $f(x)$  in the basis of the ${\mathbb J_j^{m,q}(x)}$ \eqref{22} 
%\begin{widetext}
\begin{equation}\begin{array}{lll}\label{030}
(K_+f)(x)&=& \ds\sum_j f_j \,
K_+\,\mathbb J^{m,q}_{j}(x)\\[0.4cm]
&=& \ds\sum_j f_j \,
\sqrt{j+1/2}\,
K_+\,\mathcal J^{m,q}_{j}(x)\\[0.4cm]
&=&\ds
\sum_j f_j \, \sqrt{(j+1)^2-m^2}\,\sqrt{j+1/2}\,\mathcal J^{m,q}_{j+1}(x)
\\[0.4cm]
&=&\ds \sum_j f_j \,\frac{\sqrt{(j+1)^2-m^2}\,\sqrt{j+1/2}}{\sqrt{j+3/2}}\,
\mathbb J^{m,q}_{j+1}(x)\,.
\end{array}\end{equation}
To show that $(K_+f)(x)\in\Phi_{m,q}$, we have to check that for any $r,s=0,1,2,\dots$, the following series converge:
\begin{equation}\label{031}
[p_{r,s}(K_+f)]^2= \sum_j |f_j|^2\,\frac{{j+1/2}}{{j+3/2}}\,[(j+1)^2-m^2]\,(j+1+|m|+1)^{2r}\,(j+1+|q|+1)^{2s} \,.
\end{equation}
Note that
\begin{equation}\label{032}
(j+|m|+2)^2= (j+|m|+1+1)^2 = (j+|m|+1)^2 + 2 (j+|m|+1) +1 \le 4(j+|m|+1)^2\,,
\end{equation}
where we have taken into account  that if $a$ and $b$ are two positive numbers, then, $2ab\le a^2+b^2$. Also, we have that
\begin{equation}\label{033}
0\le (j+1)^2-m^2 \le (j+1)^2+m^2 \le (j+1)^2+m^2+2|m|\,(j+1) = (j+|m|+1)^2\,,
\end{equation}
and $(j+1/2)/(j+3/2)<1$.
If we use \eqref{032} and \eqref{033} in \eqref{031}, we conclude that
\begin{equation}\label{34}
[p_{r,s}(K_+f)]^2\;\le\; 2^{2r+2s}  \sum_j|f_j|^2\,(j+|m|+1)^{2(r+1)}\,(j+|q|+1)^{2s} = 2^{2r+2s}\,[p_{r+1,s}(f)]^2 \,.
\end{equation}%\end{widetext}
This proves that $K_+f\in\Phi_{m,q}$. In addition, the inequality \eqref{34} shows the continuity of $K_+$ on $\Phi_{m,q}$, after \eqref{024}. A similar proof works for $K_-$. Thus, we have concluded that $J$, $M$, $Q$ and $K_\pm$ are continuous linear operators on all the spaces $\Phi_{m,q}$. Furthermore, since $J$, $M$, $Q$ are symmetric on $\Phi_{m,q}$, they can be continuously extended into the duals $\Phi_{m,q}^\times$ by using the duality formula \eqref{2}. Same for $K_\pm$ as they are the formal adjoint of each other, i.e., $(K_\pm)^\dagger=K_\mp$.
%%%%%%%%%%%%%%%%%%%%%
%%%%%%%%%%%%%%%%%%%%%
\subsection{Operators mapping into different spaces $\Phi_{m,q}$}\label{3C}

Let us go back to equations \eqref{11}  which define mappings between different $\Phi_{m,q}$. For instance, mappings $A_\pm$ and $B_\pm$  transform spaces with labels $m$ and $q$ integers (half-integers) into other with $m$ and $q$ also integers (half-integers). On the other hand, the remaining operators of \eqref{11} define mappings from integer (half-integer) indices into half-integer (integer) indices. In any case, these operators are well defined and continuous between their corresponding spaces. It is enough to prove this fact with $A_+$, being the proofs for the remaining cases similar. As usual, we consider the action of $A_+$ on $f(x)\in\Phi_{m,q}$ as
%\begin{widetext}
\begin{equation}\begin{array}{lll}\label{35}
(A_+f)(x) &=&\ds  A_+ \left(\sum_{j\ge \sup (|m|,|q|)} f_j\,
\mathbb J_j^{m,q}(x)\right) =\sum_{j\ge \sup (|m|,|q|)} f_j\,A_+\,
\mathbb J_j^{m,q}(x) \\[0.6cm]
&=&\ds  \sum_{j\ge \sup (|m+1|,|q|)}  f_j \sqrt{(j-m)(j+m+1)}\,
\mathbb J_j^{m+1,q}(x)\,.
\end{array}\end{equation}
Observe that, if $|q|>|m|$, the same coefficients $f_j$ must appear in both sums in \eqref{35}.
On the contrary, if $|m|\ge |q|$, the term with coefficient $f_j$ with $j=|m|$, does not appear in the second sum in \eqref{35} since $A_+\,\mathcal J_j^{j,q}(x)=0$ (see \eqref{11}). Next, let us consider the following series for any $r,s=0,1,2,\dots$
\begin{equation}\label{36}
[p_{r,s}^{m+1,q}(A_+f)]^2=\sum_{j\ge \sup\{|m+1|,|q|\}} |f_j|^2 \,[(j-m)(j+m+1)]\,(j+|m+1|+1)^{2r}\,(j+|q|+1)^{2s}\,,
\end{equation}
where $\{p_{r,s}^{m+1,q}\}_{r,s\in \N}$ is the set of seminorms in $\Phi_{m+1,q}$. Since,
\begin{equation*}\label{37}
(j+|m+1|+1)^2 \le (j+|m|+1+1)^2 \le 4\,(j+|m|+1)^2\,,
\end{equation*}
we have that
\begin{equation*}\label{038}
[p_{r,s}^{m+1,q}(A_+f)]^2\; \le \; 2^2 \sum_{j\ge \sup(|m|,|q|)} |f_j|^2 \,(j+|m|+1)^{2(r+1)}\,(j+|q|+1)^{2s} =  [2\,p_{r+1,s}^{m,q}(f)]^2\,.
\end{equation*}
Then after \eqref{024}, the linear mapping $A_+:\Phi_{m,q}\longmapsto \Phi_{m+1,q}$ is continuous. By a simple generalization of the duality formula \eqref{2}, its adjoint, $A^\dagger_+$, is a linear continuous mapping between the duals: $A_+^\dagger :\Phi_{m+1,q}^\times \longmapsto \Phi_{m,q}^\times$. Since $\Phi_{m,q}\subset \Phi_{m,q} ^\times$ for all $m$ and $q$, it is legitimate to investigate the form of the restriction of $A_+^\dagger$ to $\Phi_{m+1,q}$.%\end{widetext}

The duality formula gives for all the NAJF :
\begin{equation*}\label{039}
\langle \mathbb J_{j'}^{m+1,q}|A_+\,\mathbb J_j^{m,q}\rangle = \langle A_+^\dagger \mathbb J_{j'}^{m+1,q}|\, \mathbb J_j^{m,q}\rangle\,,
\end{equation*}
that gives the identity between two brackets. The l.h.s. of each bracket represents a functional acting on the vector in the r.h.s. of the bracket. Thus, $ \mathbb J_{j'}^{m+1,q}$ is a functional acting on the vector $A_+\,\mathbb J_j^{m,q}\in\Phi_{m+1,q}$ and $A_+^\dagger \mathbb J_{j'}^{m+1,q}$ is a functional on $\mathbb J_j^{m,q}\in\Phi_{m,q}$. By elementary properties of RHS, this action coincides with the scalar product on the Hilbert space, which in the present case is $L^2[-1,1]$. This fact shows that
%\begin{widetext}
\begin{equation}\begin{array}{lll}\label{040}
\langle \mathbb J_{j'}^{m+1,q}|A_+\,\mathbb J_j^{m,q}\rangle &=&\ds
 \sqrt{(j-m)(j+m+1)}\, \langle \mathbb J_{j'}^{m+1,q}|\mathbb J_j^{m+1,q}\rangle \\[2ex]
 &=&\ds  \sqrt{(j-m)(j+m+1)}\, \delta_{j,j'} \\[2ex]
 &=&\ds \langle \sqrt{(j-m)(j+m+1)}\, \mathbb J_j^{m,q}|\mathbb J_{j'}^{m,q}\rangle = \langle A_-\,\mathbb J_j^{m+1,q}|\mathbb J_{j'}^{m,q}\rangle\,,
\end{array}\end{equation}\end{widetext}
where the last identity in \eqref{040} comes from the action of the operator $A_-$ on the $\mathcal J_j^{m,q}$ after \eqref{11}, i.e.
\begin{equation*}\label{41}
A_-\,\mathbb J_j^{m+1,q} = \sqrt{(j+m+1)(j-m)}\, \mathbb J_j^{m,q}\,.
\end{equation*}
Therefore, $A_-=A_+^\dagger$ on $\Phi_{m+1,q}$. Since $\Phi_{m+1,q}$ is dense in its dual $\Phi_{m+1,q}^\times$ and the properties of RHS, $A_-$  can be uniquely extended to a continuous linear mapping from $\Phi_{m+1,q}^\times$ to $\Phi_{m,q}^\times$. This extension coincides with $A_+^\dagger$.

A similar discussion may be applied to all other mappings \eqref{11}. In each case, it is obvious which are the involved spaces and which topologies we should endow on each of these spaces.

%%%%%%%%%%%%%%%%%%%%%%%%%%%%%%%%%%%%%%%%%%%%%%%%%%%%%%%%%
%%%%%%%%%%%%%%%%%%%%%%%%%%%%%%%%%%%%%%%%%%%%%%%%%%%%%%%%%
\section{%Discrete and continuous b
Bases with $m$ and $q$ fixed}\label{discretecontinuousbases}

So far, we have considered realizations of RHS of the form
\[
\Phi_{m,q}\subset  L^2[-1,1] \subset \Phi_{m,q}^\times\,, \quad \forall \, m,q\in \Z\,\,(\text{or}\; \Z+1/2)
\]
 where  the elements of the test spaces $\Phi_{m,q}$ are functions.

 We also may consider an abstract RHS
  $\Psi_{m,q}\subset \mathcal H_{m,q} \subset \Psi_{m,q}^\times$, where the elements of each of the spaces are just {\it vectors} that we obtain by means of an unitary operator  
  \begin{equation}\label{420}
  U:\mathcal H _{m,q} \longmapsto L^2[-1,1] \,,
  \end{equation}
   such that $U\Psi_{m,q}=\Phi_{m,q}$ and $U$ is one-to-one and onto topology preserving mapping.  
%  Let us consider another RHS $\Phi\subset \mathcal G\subset\Phi^\times$. We say that these RHS are unitarily equivalent if there exists a unitary operator $U:\mathcal H \longmapsto\mathcal G$, such that $U\Psi_{m,q} =\Phi_{m,q} $ and $U$ is one-to-one and onto topology preserving mapping. 
Then, $U^\dagger\Phi_{m,q} =U^{-1}\Phi_{m,q} =\Psi_{m,q} $ has the same properties. With the aid of the duality formula similar to \eqref{2},
\begin{equation*}\label{42}
\langle F|U^\dagger\varphi\rangle = \langle UF|\varphi\rangle\,, \qquad \forall\,\varphi\in\Psi_{m,q} \,, \;\forall\,F\in \Psi_{m,q} ^\times\,,
\end{equation*}
we may extend $U$ to a linear continuous one-to-one mapping  $U \Psi_{m,q}^\times=\Phi_{m,q}^\times$, preserving topologies.
In other words, we have the following  diagram
 \[
\begin{array}{lllclll}
&\Psi_{m,q} &\subset &{\mathcal H_{m,q} }&\subset&\Psi_{m,q} ^\times
\\[0.2cm]
  & \hskip-0.25cm  {U}\downarrow && \hskip-0.30cm {U}\downarrow &&\hskip-0.30cm {U}\downarrow
\\[0.2cm]
&\Phi _{m,q} &\subset  &L^2[-1,1] &\subset   &\Phi_{m,q} ^\times
\end{array}\,.
\]
%If the Hilbert space $\mathcal G$ is a space of functions, we say that $\Phi\subset \mathcal G\subset\Phi^\times$ is a representation of $\Psi\subset \mathcal H \subset \Psi^\times$ and that  $\Phi$ is a space of {\it test functions}. This is the case when $\mathcal G \equiv L^2[-1,1]$. Then, we have a unitary mapping $U:\mathcal H\longmapsto L^2[-1,1]$. As far as $\mathcal H$ is abstract, then $U$ is arbitrary.
%%%%%%%%%%%%%%%%%%%%%%%%
%%%%%%%%%%%%%%%%%%%%%%%%
\subsection{Discrete basis}

For the forthcoming discussion, we need to introduce some new ingredients. These are some new RHS to be defined later. In any case, for each of the values of $m$ and $q$, we shall define in Section~\ref{cuatroc} a space of test functions $\Xi_{m,q}$ that include all the NAJF $\mathbb J_j^{m,q}(x)$ with $j\ge\sup (|m|,|q|)$. The spaces $\Xi_{m,q}$ will be different from the spaces $\Phi_{m,q}$, although they will play a similar role.

Then for fixed $m$ and $q$, we take $\Omega_{m,q}:=U^{-1}\,\Xi_{m,q}$, as well as $\widetilde K_\pm := U^{-1}\,K_\pm\,U$ and
\begin{widetext}\begin{equation}\label{jmq}
|j,(m,q)\rangle:= U^{-1}\mathbb J_j^{m,q} = U^{-1}\,\sqrt{j+1/2}\,\mathcal J_j^{m,q}(x)
= \sqrt{j+1/2}\;U^{-1}\;\mathcal J_j^{m,q}(x)\,.
\end{equation}
 After  \eqref{030}, we have that
 \begin{equation}\begin{array}{lll}\label{43}
\widetilde K_+|j,(m,q)\rangle &=&\ds\frac{\sqrt{(j+1)^2-m^2}\,\sqrt{j+1/2}}{\sqrt{j+3/2}}\,|j+1,(m,q)\rangle\,,\\[0.4cm]
 \widetilde K_-|j,(m,q)\rangle& =&\ds\frac{ \sqrt{j^2-m^2}\,\sqrt{j+1/2}}{\sqrt{j-1/2}}\,\,|j-1,(m,q)\rangle\,,
\end{array}\end{equation}
with $\widetilde K_-\,|0, (0,0)\rangle=0$. For $m$ and $q$ fixed and $j\ge\sup\{|m|,|q|\}$, the vectors $|j,(m,q)\rangle$ form a basis in $\mathcal H$, so that 
\begin{equation}\label{44}
\langle j,(m,q)|j',(m,q)\rangle =\delta_{j,j'}\,,\qquad \sum_j |j,(m,q)\rangle\langle j,(m,q)| = \mathcal I\,,
\end{equation}
where $\mathcal I$ is the identity on $\mathcal H_{m,q}$ and unless otherwise stated, sum on $j$ will run from $j\ge\sup\{|m|,|q|\}$ to infinite.

%%%%%%%%%%%%%%%%%%%%%%%%
%%%%%%%%%%%%%%%%%%%%%%%%
\subsection{Continuous basis}

Next, we {\it formally} construct a {\it continuous basis} for $\Phi_{m,q}$ as follows: for any $x\in[-1,1]$, we define
\begin{equation}\label{45}
|x,(m,q)\rangle := \sum_j |j,(m,q)\rangle \,
\mathbb J_j^{m,q}(x)\,.
\end{equation}
This formal definition, along standard manipulations give some properties for kets $|x,(m,q)\rangle$ such as \cite{COV}
\begin{equation}\label{46}
\langle x,(m,q)|x',(m,q)\rangle = \delta(x-x')\,, \qquad \int_{-1}^1 |x,(m,q)\rangle\,dx\,\langle x,(m,q)|=I\,,
\end{equation}
where $I$ is some sort of identity. In addition,
\begin{equation}\label{47}
|j,(m,q)\rangle =\int_{-1}^1 |x,(m,q)\rangle \,
\mathbb J_j^{m,q}(x)\,dx\,,
\end{equation}
and
\begin{equation}\label{48}
\mathbb J_j^{m,q}(x) =
\langle x,(m,q)|j,(m,q)\rangle= 
 \langle j,(m,q)|x,(m,q)\rangle\,.
\end{equation}
Note that the $\{ \mathbb J_j^{m,q}(x)\}$ are the (real) transition matrices between the discrete and continuous bases of elements 
$|j,(m,q)\rangle$ and  $|x,(m,q)\rangle$.

For any arbitrary $|f\rangle \in \Phi_{m,q}$, we have two different expressions depending the basis where it is spanned
\begin{equation}\label{49}
f\equiv |f\rangle=\left\{
\begin{array}{lll}\ds
  \int_{-1}^1 dx\, |x,(m,q)\rangle \,f^{m,q}(x)\\[0.4cm]
 \ds \sum_{j\ge\sup(|m|,|q|)} f^{m,q}_j \,|j,(m,q)\rangle
  \end{array}\right.
\end{equation}
with
\begin{equation}\label{50}
\begin{array}{lllll}\ds
f^{m,q}(x)&=& \ds\langle x,(m,q)|f\rangle &=&\ds \sum_j \
f^{m,q}_j\, \mathbb J^{m,q}_j(x)\,,\\[0.4cm]
f^{m,q}_j &=&\ds \langle j,(m,q)|f\rangle &=&\ds \int_{-1}^1 
\mathbb J_j^{m,q}(x) \,f^{m,q}(x)\,dx\,.
\end{array}\end{equation}
Thus, we have obtained a relation between discrete, $\{|j,(m,q)\rangle\}_{j\ge\sup(|m|,|q|)}$, and continuous basis,
$\{|x,(m,q)\rangle\}_{x\in [-1,1]}$.

%%%%%%%%%%%%%%%%%%%%%%%%
%%%%%%%%%%%%%%%%%%%%%%%%
\subsection{The spaces of test functions $\Xi_{m,q}$}\label{cuatroc}

The next goal is to provide a good definition of the kets $|x,(m,q)\rangle$. This is why we need a new RHS with spaces of test functions denoted by $\Xi_{m,q}$. These functions should have the following properties:

\begin{enumerate}
\item Each function $f(x)\in\Xi_{m,q}$ is of the form \eqref{22}.
\item For any $r,s=0,1,2,\dots$, we have that the coefficients in the series \eqref{22} satisfy
\begin{equation}\label{52}
t^{m,q}_{r,s}(f):= \sum_{{j\ge \sup(|m|,|q|)}}^\infty |a_j|\,(j+|m|+1)^r\,(j+|q|+1)^s<\infty\,.
\end{equation}
\end{enumerate}
Relation \eqref{52} implies that  the series \eqref{22} converge in the Hilbert space norm as well as {\it pointwise} and {\it absolutely}. For convenience in the presentation, we shall prove this properties later. 

In the sequel, we shall denote the mappings $t^{m,q}_{r,s}$ as $t_{r,s}$ unless otherwise stated to avoid confusion. These,  $t_{r,s}$ (for $r,s=0,1,2,\dots$) are a family of norms on $\Xi_{m,q}$ that define the topology on $\Xi_{m,q}$.  To complete a RHS, one has to add $L^2[-1,1]$ as Hilbert space. The spaces $\Xi_{m,q}$ are dense in $L^2[-1,1]$ since they contain all finite linear combinations of $\mathbb J_j^{m,q}(x)$. However, we need more information to conclude that $\Xi_{m,q}\subset L^2[-1,1] \subset \Xi_{m,q}^\times$ are RHS. Here, we define $\Xi_{m,q}^\times$ as usual and note that this dual space is the same no matter whether $\Xi_{m,q}$ is complete or not. This extra information is giving by showing that the canonical injection
\[\begin{array}{lcl}
\mathfrak j:&\Xi_{m,q}\;\;&\longmapsto \;\;L^2[-1,1]\\[0.4cm]
&\hskip-0.2cm f\;\;&\longmapsto \;\;\hskip0.75cm \mathfrak j(f)
\end{array}\]
 is continuous. This proof is very simple:
since
\begin{equation}\label{53}
||\mathfrak j(f)|| =\sqrt{\sum_{j\ge \sup(|m|,|q|)} |f_j|^2} \le \sum_{j\ge \sup(|m|,|q|)} |f_j|=t_{0,0}(f)\,,
\end{equation}\end{widetext}
where $||-||$ denotes the Hilbert norm on $L^2[-1,1]$. Then, the conclusion comes from a simple generalization of \eqref{024}. Thus,
\[
\Xi_{m,q}\subset L^2[-1,1] \subset \Xi_{m,q}^\times
\]
 is a RHS. 
 
 Now, combine \eqref{52} with $r=s=0$ along to the inequality in \eqref{53} to prove our assertion that the series in \eqref{22} satisfying \eqref{52} converges in the Hilbert space norm. Pointwise and absolute convergences will be a consequence of the Weisstrass M-theorem  \cite{weber-arfken2004} as shall see later.

 %%%%%%%%%%%%%%%%%%%%%%%%
%%%%%%%%%%%%%%%%%%%%%%%%
\subsection{The spaces of test vectors $\Omega_{m,q}$} 

Let us consider the unitary operator $U$ introduced in  \eqref{420} and define the spaces $\Omega_{m,q}:=U^{-1}\,\Xi_{m,q}$, for each pair of values $m,q$. Let us transport the topology defined on the space of test functions $\Xi_{m,q}$ to the space of test vectors $\Omega_{m,q}$ by means of $U^{-1}$. It comes that for any $|f\rangle=\sum_j f_j^{m,q}\,|j,(m,q)\rangle\in\O_{m,q}$, the norms defining the topology on $\O_{m,q}$ are equal to $t_{r,s}(f)$ defined as in \eqref{52}, see also the line below \eqref{52}. For fixed $m$ and $q$, the triplet $\Xi_{m,q}\subset L^2[-1,1] \subset \Xi_{m,q}^\times$ is a realization of the abstract RHS  $\Omega_{m,q}\subset \mathcal H \subset \Omega_{m,q}^\times$ by $U$. In other words, 
\[
\begin{array}{lllclll}
&\O_{m,q}&\subset &{\mathcal H}&\subset& \O_{m,q}^\times
\\[0.2cm]
  & \hskip-0.25cm  {U}\downarrow && \hskip-0.30cm {U}\downarrow &&\hskip-0.17cm {U}\downarrow
\\[0.2cm]
&\Xi_{m,q}&\subset  & L^2[-1,1] &\subset   & \Xi_{m,q}^\times
\end{array}\,.
\]

Let us discuss the validity of formula \eqref{45} to define the kets $|x,(m,q)\rangle$. To do it, we have to take into account the following inequality valid for the Algebraic Jacobi Functions   (see Appendix~C):
 \begin{widetext} \begin{equation}\label{54}
 |\mathcal J^{m,q}_j(x)|\leq (j+|m|+1)\,(j+|q|+1) \,, 
 \qquad \forall j \ge |m|, |q|;\; 2j\in\mathbb N;\; j-m,\,   j-q\in\mathbb N\,,
\end{equation}
that  for the NAJF  %$\mathbb J^{m,q}_j(x)$
 \eqref{NAJF} 
%the above inequality \eqref{54} 
becomes
\begin{equation}\label{540}
|\mathbb J^{m,q}_j(x)| \le  \sqrt{j+1/2}\,(j+|m|+1)(j+|q|+1)\,.
\end{equation}

Next, let us consider  
 $|f\rangle \in\O_{m,q}$. From  equations \eqref{45}, \eqref{49}, \eqref{50} and \eqref{NAJF} we have that 
 \begin{equation}\label{55}
f^{m,q}(x):=\langle x,(m,q)|f\rangle = \sum_{j} \langle j,(m,q)|f\rangle\,
\mathbb J_j^{m,q}(x)\,,
\end{equation}
where from now on  the sums in $j$ are always extended from $\sup(|m|,|q|)$ up to infinite, unless otherwise stated. 

Condition \eqref{52} implies the absolute and pointwise convergence of the series \eqref{22}. Indeed, let us consider
\begin{equation}\begin{array}{lll}\label{551}
\sum_j |f_j|\,|\mathbb J^{m,q}_j(x)|& \le &\ds\sum_j |f_j|\,\sqrt{j+1/2}\,(j+|m|+1)(j+|q|+1) \\[2ex]
& \le &\ds\sum_j |f_j|\, (j+|m|+1)^2(j+|q|+1)\,,
\end{array}\end{equation}
which converges by condition \eqref{52}. The first inequality in \eqref{551} is due to \eqref{540}, while the second one is trivial. Then, the Weierstrass M-theorem shows our assertion.

Since the convergence is now pointwise after our hypothesis, we may take modulus in both sides of \eqref{55} and remembering \eqref{52} so as to obtain the following inequality:
\begin{equation}\begin{array}{lll}\label{56}
\big| \langle f|x,(m,q)\rangle\big| &\le &\ds \sum_{j} |\langle f\big | j,(m,q)\rangle|\,
\cdot\big|\mathbb J_j^{m,q}(x)\big| \\[0.4cm]
 & \le &\ds  \sum_{j} \big| f_{j}|\, \sqrt{j+1/2}\, (j+|m|+1)(j+|q|+1)  \\[0.4cm]
 &\le& \ds \sum_j |f_j| \, (j+|m|+1)^2(j+|q|+1) = t_{2,1}(f)\,.
\end{array}\end{equation}

The inequality \eqref{56} along the linearity of the functional $|x,(m,q)\rangle$ on $\Omega_{m,q}$ prove the continuity of $|x,(m,q)\rangle$ on $\Omega_{m,q}$, and hence that $|x,(m,q)\rangle \in \Omega_{m,q}^\times$. 
 Note that, while $|f\rangle \in\O_{m,q}$, its realization as function is given by $\langle f|x,(m,q)\rangle$, which is in $\Xi_{m,q}$.

Next, let us prove the continuity of $\widetilde K_\pm$ on $\O_{m,q}$. For arbitrary $|f\rangle =\sum_j f_j\,|j,(m,q)\rangle \in \O_{m,q}$, we shall prove in just one operation that $\widetilde K_\pm\,|f\rangle \in\O_{m,q}$ and that
$\widetilde K_\pm$ are continuous on $\O_{m,q}$. By definition,
the action of $\widetilde K_\pm$ on $\O_{m,q}$ is given by
\[
\widetilde K_\pm \,|f\rangle:= \sum_j f_j \,
\widetilde K_\pm \,|j,(m,q)\rangle\,,\qquad \forall |f\rangle\in\O_{m,q}\,.
\]
  After the action of $\widetilde K_\pm$ on the basis vectors given in \eqref{43}, we have for $\widetilde K_+$ that
(see \eqref{23})
\begin{equation*}\label{57}
t_{r,s}(K_+\,|f\rangle) = \sum_j |f_j|\, \sqrt{(j+1)^2-m^2}\,(j+1+|m|+1)^r (j+1+|q|+1)^s\,,
\end{equation*}
for any $r,s,=0,1,2,\dots$. Then, and taking into account that
\begin{equation*}\label{58}
(j+1)^2-m^2\le (j+|m|+1)^2 \quad {\rm and} \quad (j+1+|m|+1) \le 2(j+|m|+1)\,,
\end{equation*}
we get that 
\begin{equation*}\label{59}
t_{r,s}(K_+\,|f\rangle)\leq 2^{r+s} \sum_j |f_j|\,(j+|m|+1)^{r+2} (j+|q|+1)^{s+1} = 2^{r+s}  \, t_{r+2,s+1}(f)\,,
\end{equation*}
or, in other words,
\begin{equation*}\label{60}
t_{r,s}(K_+\,|f\rangle) \le  2^{r+s}  \, t_{r+2,s+1}(f)\,,\qquad \forall\, |f\rangle\in\O_{m,q}\,.
\end{equation*}
This proves our claim for $\widetilde K_+$. A similar proof will show that $\widetilde K_-$ satisfies the  same properties. In particular, these operators can be continuously extended into the dual ${\O_{m,q}}^\times$. The same is true for $K_\pm$ acting on $\Xi_{m,q}$.

It is worthy to notice that  as a consequence of the fact that $\ds\left({\sum |f_n|^2}\right)^{1/2} \le \sum
 |f_n|$ (see \eqref{53})
one has that
\[
p_{r,s}(f)\le t_{r,s}(f)\,, \qquad \forall |f\rangle\in\O_{m,q}\,,
\]
where $p_{r,s}$ were defined in \eqref{36} and
$t^{m,q}_{r,s}$ in \eqref{52}. These properties have interesting consequences: first of all $\Xi_{m,q}\subset \Phi_{m,q}$ and second that the canonical injection $\mathfrak j:\Xi_{m,q} \longmapsto \Phi_{m,q}$, $\mathfrak j(f):=f$, is continuous. Thus, we have the relations:
\begin{equation*}\label{61}
\Xi_{m,q}\subset \Phi_{m,q} \subset L^2[-1,1] \subset \Phi_{m,q}^\times \subset \Xi_{m,q}^\times\,,
\end{equation*}
where all canonical injections (identities) are continuous. Same properties are valid for the chain of abstract spaces:
\begin{equation*}\label{62}
\O_{m,q} \subset \Psi_{m,q} \subset \mathcal H \subset \Psi_{m,q}^\times \subset \O_{m,q}^\times\,.
\end{equation*}

Thus, the topology given by the norms $t_{r,s}$ makes the linear functionals $|x,(m,q)\rangle$ on $\Omega_{m,q}$ continuous, a property that may not be true if instead we would have used the topology produced by the norms $p_{r,s}$ defined on each of the spaces $\Psi_{m,q}$.

We finish this section by adding that the operators $J$, $M$ and $Q$ as defined in \eqref{jmq} are also continuous on all the spaces $\Xi_{m,q}$. The proof is immediate. Also there are some relations between spaces which are also trivially continuous. For instance:
\begin{equation*}\label{63}
\CD
\Xi_{m,q} @>\mathfrak j >> \Phi_{m,q}  @> A_+ >> \Phi^{m+1,q}\,,
\endCD
\end{equation*}
where $\mathfrak j$ is the identity. Similar relations can be constructed with the operators defined in the list \eqref{11}.
%%%%%%%%%%%%%%%%%%%%%%%%%%%%%%%%%%%%%%%%%%%%%%%%%%%%%%
%%%%%%%%%%%%%%%%%%%%%%%%%%%%%%%%%%%%%%%%%%%%%%%%%%%%%%

\section{Jacobi Harmonics and Jacobi Fermion Harmonics}\label{totalspaces}

The Associated Legendre polynomials 
 (that depend on two parameters and one variable) are related to the Spherical Harmonics (two parameters and two variables).
 Similarly the Algebraic Jacobi Functions, that depend on three parameters and one variable, can be related  to the ``Jacobi Harmonics''
\begin{equation}\label{5.1}
\mathcal N^{m,q}_j(x,\phi,\chi) :=\mathbb J^{m,q}_j(x)\,e^{im\phi}\,e^{iq\chi}= \sqrt{j+1/2}\,\mathcal J^{m,q}_j(x)\,e^{im\phi}\,e^{iq\chi}\,, 
\end{equation}%\end{widetext}
with  three discrete parameters $(j,m,q)$ and three continuous variables  $(x ,\,\phi,\chi )$.
%where $\phi$ and $\chi$ are two angular variables. 

  We have to distinguish two cases depending if $(j,m,q)$ are integer or half-integer. If they are integer   the range of the angular variables is  $\phi\in[0,2\pi)$ and $\chi\in[0,\pi]$, and we call $\mathcal N^{m,q}_j(x,\phi,\chi)$ properly Jacobi Harmonics  while if they are half-integer  $\phi\in[0,4\pi)$ and $\chi\in[0,2\pi)$  and we call them  ``Jacobi Fermion Harmonics''. 

Indeed,  the Jacobi Harmonics are a  generalization of the  Spherical Harmonics: assuming $x:=\cos\theta$ we have $\theta \in[0,\pi]$  and  the  Jacobi Harmonics are defined in the hypersphere   $S^3$, and satisfy the relations of orthogonality  and completeness 
%\begin{widetext}
\begin{equation}\begin{array}{rll}\label{5.2}
\ds\frac 1{2\pi^2} \int_0^{2\pi} d\phi \int_0^{\pi} d\chi \int_{-1}^1 dx \;\mathcal N^{m,q}_j(x,\phi,\chi)\,\left({\mathcal N^{m',q'}_{j'}}\right)^*(x,\phi,\chi) &=&\delta_{j,j'}\,\delta_{m,m'}\,\delta_{q,q'}\,,\\[0.5cm]
\ds\sum_{j=0}^{\infty}\,\sum_{m,q=-j}^j \left[\mathcal N^{m,q}_j(x,\phi,\chi)\right]^*\, \mathcal N^{m,q}_j(x',\phi',\chi')&=&\delta(x-x')\,\delta(\phi-\phi')\,\delta(\chi-\chi')\,.
\end{array}
\end{equation}
The Jacobi Harmonics thus   behave on $S^3$ like the Spherical Harmonics on $S^2$ \cite{CGO19}. On the other hand, for $j$  half-integer  the Jacobi Fermion Harmonics  are  the generalization of the Fermionic quasi-Spherical Harmonics  \cite{hunter99} because they verify 
\begin{equation}\begin{array}{rll}\label{5.3}
\ds \frac 1{8\pi^2} \int_0^{4\pi} d\phi \int_0^{2\pi} d\chi \int_{-1}^1 dx \;\mathcal N^{m,q}_j(x,\phi,\chi)\,\left({\mathcal N^{m',q'}_{j'}}\right)^*(x,\phi,\chi) &=&\delta_{j,j'}\,\delta_{m,m'}\,\delta_{q,q'}\,,
\\[0.5cm]
\ds\sum_{j=1/2}^{\infty}\,\sum_{m,q}  \left[\mathcal N^{m,q}_j(x,\phi,\chi)\right]^*\, \mathcal N^{m,q}_j(x',\phi',\chi')&=&\delta(x-x')\,\delta(\phi-\phi')\,\delta(\chi-\chi')\,.
\end{array}\end{equation}

For any fixed $j$ no matter whether integer or half-integer, the functions $e^{im\phi}$ and $e^{iq\chi}$ span respective vector spaces of dimension $2j+1$, which are isomorphic to $\mathbb C^{2j+1}$. Therefore, we may identify these spaces of functions with  $\mathbb C^{2j+1}$, and  establish the following result:
\medskip

{\bf Theorem}.- {\it %For $j$ either integer or half-integer, t
The set of functions $\mathcal N_j^{m,q}(x,\phi,\chi)$ with $m,q=-j,-j+1,\dots,j-1,j$ and  $j=0,1,2,\dots$, or $j=1/2,3/2\dots$,  is a basis of the Hilbert spaces:
\begin{equation*}\label{5.4}
\mathcal L_H= \bigoplus_{j=0}^\infty L^2[-1,1]\otimes \mathbb C^{2j+1}\otimes \mathbb C^{2j+1}
\,,\qquad \mathcal L_F:=\bigoplus_{j=1/2}^\infty L^2[-1,1]\otimes \mathbb C^{2j+1}\otimes \mathbb C^{2j+1}
\,.
\end{equation*}
 The subindices $H$ and $F$ stand for ``Jacobi Harmonic'' and ``Jacobi Fermion Harmonic'', respectively}.
\medskip

{\bf Proof}. It is a direct outcome of \eqref{5.2} and \eqref{5.3}. \hfill$\blacksquare$
\medskip

As a consequence, functions  $f_H(x,\phi,\chi)\in \mathcal L_H$ and in $f_F(x,\phi,\chi)\in \mathcal L_F$ admit the following expansions, which converge in the Hilbert norm:
\begin{eqnarray}
  f_H(x,\phi,\chi) &=&  \sum_{j=0}^\infty \;\sum_{m,q=-j}^{j} f^{(H)j}_{m,q}\,  \mathcal N_j^{m,q}(x,\phi,\chi)\,,\hskip0.7cm
  ||f_H(x,\phi,\chi)||^2 =\sum_{j,m,q}  \big| f^{(H)j}_{m,q}\big|^2 <\infty\,, \label{5.5} \\[2ex]
  f_F(x,\phi,\chi) &=&\displaystyle \sum_{j=1/2}^\infty\; \sum_{m,q=-j}^{j}  f^{(F)j}_{m,q}\, \mathcal N_j^{m,q}(x,\phi,\chi)\,,\quad\ds
  ||f_F(x,\phi,\chi)||^2 = \sum_{j,m,q}  \big| f^{(F)j}_{m,q}\big|^2 <\infty\label{5.6}\,.
\end{eqnarray}

Let $\Phi_H\subset \mathcal L_H$ and $\Phi_F\subset\mathcal L_F$ be the spaces of functions $f_H(x,\phi,\chi)$ and $f_F(x,\phi,\chi)$ as in \eqref{5.5} and \eqref{5.6}, respectively, %, such that  for all the admissible values of $j$, $m$ and $q$ and 
that for all $r,s=0,1,2,\dots$, verify
\begin{eqnarray*}
[p^{H}_{r,s}(f_H)]^2:=\sum_{j=0}^\infty \;\sum_{m,q=-j} ^j \big| f^{(H)j}_{m,q}\big|^2\,(j+|m|+1)^{2r}\,(j+|q|+1)^{2s} <\infty\,, \label{5.7}\\[2ex]
[p^{F}_{r,s}(f_F)]^2:=\sum_{j=1/2}^\infty \;\sum_{m,q=-j}  ^j \big| f^{(F)j}_{m,q}\big|^2\, (j+|m|+1)^{2r}\,(j+|q|+1)^{2s} <\infty\,. \label{5.8}
\end{eqnarray*}
For each $r,s=0,1,2,\dots$, $p^{H,F}_{r,s}(f_{H,F})$ are norms for the function $f_{H,F}$. With the topology produced by these norms, $\Phi_H$ and $\Phi_F$ are Fr\'echet nuclear  spaces. In addition,  $\Phi_H$ and $\Phi_F$ contain the functions in the respective basis of $\mathcal H_H$ and $\mathcal H_F$, so that they are dense in the corresponding Hilbert space. Furthermore,  for $r=s=0$, we obtain the Hilbert norm, so that the canonical injections $\mathfrak j_{H,F}:\Phi_{H,F}\longmapsto \mathcal L_{H,F}$ are continuous. All these facts imply that
\[
\Phi_{H,F}\subset \mathcal L_{H,F} \subset  \Phi_{H,F}^\times
\]
and also
\[
\Phi_{H\oplus F}\subset \mathcal L_{H\oplus F} \subset \Phi_{H\oplus F}^\times \,,
\]
where $\Phi_{H\oplus F}=\Phi_{H}\oplus \Phi_{ F}$ and  $\mathcal L_{H\oplus F}=\mathcal L_{H}\oplus \mathcal L_{ F}$, are all rigged Hilbert spaces.  
%%%%%%%%%%%%%%%%%%%
%%%%%%%%%%%%%%%%%%%%
\subsection{Operators acting on $\mathcal L_{H\oplus F}$}\label{5.A}

The  set of Jacobi functions $\left\{\mathcal N^{m,q}_j(x,\phi,\chi)\right\}_{j,m,q \in N/2}$ supports a representation  of $SU(2,2)$, that  contrarily to what happens for the $\mathbb J^{m,q}_j(x)$, is also unitary and irreducible. 
Its generators are related  with those of \eqref{11} 
by
\begin{equation}\begin{array}{llllll}\label{011}
\mathcal A_\pm &=& A_\pm \, e^{\pm i\phi }\,,\qquad 
&\mathcal B_\pm &=& B_\pm \,e^{\pm i\chi }\,,  \\[0.4cm]
\mathcal C_\pm  &=& C _\pm \, e^{\pm i\phi /2}\, e^{\pm i\chi /2}\,, \qquad 
&\mathcal D_\pm  &=& D _\pm \, e^{\pm i\phi /2}\, e^{\mp i\chi /2}\,, \\[0.4cm]
\mathcal E_\pm  &=& E _\pm \, e^{\mp i\phi /2}\, e^{\pm i\chi /2}\,, \qquad 
&\mathcal F_\pm &=& F _\pm \, e^{\mp i\phi /2}\, e^{\mp i\chi /2}\,,
\end{array}\end{equation}
while operators \eqref{jmqoperators} remain invariant, i.e,  $
\mathcal J =J,\; \mathcal M=M,\;\mathcal Q=Q$.
Moreover, for $(m,q)$ fixed the functions $\mathcal N^{m,q}_j(x,\phi,\chi)$ support  a UIR of   $SU(1,1)$ generated by  the same operators \eqref{su11}, i.e., 
$\mathcal K_\pm =K_\pm,\; \mathcal K_3=K_3$.

All the above mentioned operators are continuous as we prove at the following. The proof of the continuity of $\mathcal J$
on  $\Phi_H$ (and on $\Phi_F$) is similar to that  given in \eqref{026} and  \eqref{027} for $J$.  For  $\mathcal M$ on  $\Phi_H$  
we have
\begin{equation*}\label{5.9}
(\mathcal M f_H)(x,\phi,\chi) := \sum_{j=0}^\infty \sum_{m,q} m \, f^{(H)j}_{m,q} \,\mathcal N_j^{m,q}(x,\phi,\chi)\,, \quad\qquad 
\forall f_H(x,\phi,\chi)\in \Phi_H\,.
\end{equation*}
Then,
\begin{equation*}\begin{array}{lll}\label{5.10}
[p^H_{r,s}(\mathcal  M f_H)]^2 &=&  \ds\sum_{j=0}^\infty \sum_{m,q}  \big| f^{(H)j}_{m,q}\big|^2 \,m^2 \,(j+|m|+1)^{2r}\,(j+|q|+1)^{2s}
\\[2ex] &\le & \ds\sum_{j=0}^\infty \sum_{m,q}  \big| f^{(H)j}_{m,q}\big|^2 \, (j+|m|+1)^{2(r+1)}\,(j+|q|+1)^{2s}
=[p^H_{r+1,s}(f_H)]^2\,,
\end{array}\end{equation*}
which proves that $\mathcal M\,\Phi_H\subset \Phi_H$ with continuity. Same proof is similar for $\mathcal  M$ on $\Phi_F$ and for  $\mathcal Q$ on both spaces. A consequence is that we can extend $\mathcal J$, $\mathcal M$ and $\mathcal Q$ to the duals $\left( \Phi_{H,F}\right)^\times$ with continuity with respect to the weak topology.

The operators $\mathcal A_\pm$   are reduced by the spaces $\Phi_{H,F}$ (i.e.,  $\mathcal A_\pm \Phi_{H,F}\subset \Phi_{H,F}$)  as well as  $\mathcal B_\pm$ and  $\mathcal K_\pm$. For 
$\mathcal A_+$ on  $\Phi_H$  
 we should define the action of $\mathcal A_+$ as
\begin{equation*}\label{5.11}
(\mathcal A_+\,f_H)(x,\phi,\chi):= \sum_{j=0}^\infty \sum_{m,q}  f^{(H)j}_{m,q} \sqrt{(j-m)(j+m+1)}\,\mathcal N_j^{m+1,q}(x,\phi,\chi)\,.
\end{equation*}
To prove that $(\mathcal  A_+\,f_H)(x,\phi,\chi)$ is in $\Phi_H$, we have to show that the series
\begin{equation}\label{5.12}
\sum_{j=0}^\infty\sum_{m,q} \big| f^{(H)j}_{m,q}\big|^2\,  |(j-m)(j+m+1) |\, (j+|m+1|+1)^{2r}\,(j+|q|+1)^{2s}
\end{equation}
converges for $r,s=0,1,2,\dots$. Clearly, $j+|m+1|+1 \le 2(j+|m|+1)$, so that \eqref{5.12} is smaller or equal to
\begin{equation*}\label{5.13}
2^{2r} \sum_{j,m,q} \big| f^{(H)j}_{m,q}\big|^2 (j+|m|+1)^{2(r+1)}(j+|q|+1)^{2s}\,,
\end{equation*}
which converges for all $r,s=0,1,2,\dots$. This shows that, for all $f_H(x,\phi,\chi)\in\Phi_H$, we have that  
$(\mathcal A_+\,f_H)(x,\phi,\chi)\in \Phi_H$. Moreover, this discussion also concludes that
\begin{equation*}\label{5.14}
[p^H_{r,s}(\mathcal A_+\,f_H)]^2 \le 2^{2r}\,  [p^H_{r+1,s}(f_H)]^2\,,\qquad \forall f_H\in\Phi_H\,,
\end{equation*}
which, after \eqref{024}, shows the continuity of $\mathcal A_+$ on $\Phi_H$. A similar argument is valid on $\Phi_F$ and the same considerations are in order for $\mathcal A_-$, $\mathcal B_\pm$ and   $\mathcal K_\pm$ on  $\Phi_H$ and  
$\Phi_F$.
%\end{widetext}

%The discussion at the end of  Subsection~\ref{3C}  can be weakly extended to a continuous linear mapping
%on $(\Phi_{H,F})^times : \cal A_+^\dagger (\Phi_{H,F})^\times \in (\Phi_{H,F})^times$ and analogously for
%$\cal A_-^\dagger, \cal B_\pm^\dagger, \cal K_\pm^\dagge$"
% can be weakly extended to a continuous linear mapping
%on $(\Phi_{H,F})^x: \cal A_+^\dagger (\Phi_{H,F})^x \in (\Phi_{H,F})^x$ and analogously for
%$\cal A_-^\dagger, \cal B_\pm^\dagger, \cal K_\pm^\dagger$

Also the discussion at the end Subsection~\ref{3C}  referred to the adjoint operator $A_+ ^\dagger$  can be repeated now  for $ \Phi_{H}$ and $\Phi_{F}$. Thus, $\mathcal A_+^\dagger=\mathcal A_-$ can be weakly extended to a continuous linear mapping
on $(\Phi_{H,F})^\times$  such that $\mathcal A_+^\dagger \Phi_{H,F}^\times \subset \Phi_{H,F}^\times$, and analogously for
$\cal A_-^\dagger, \cal B_\pm^\dagger, \cal K_\pm^\dagger$.

The remaining  operators of \eqref{011} ($\mathcal C_\pm,\,\mathcal D_\pm,\,\mathcal E_\pm,\, \mathcal F_\pm$) have a slight different nature. In fact, they transform  Jacobi Harmonics  into Jacobi Fermion Harmonics and vice-versa, as may be seen from \eqref{11} and \eqref{011}. As a simple consequence, these operators are mappings from $\Phi_H$ into $\Phi_F$ and vice-versa. The proof of the continuity of all these operators on these spaces can be made by similar arguments to those used so far for $\mathcal A_+$. 

Moreover they are  formal adjoint of each other, i.e., for  $\mathcal C_+$ and $\mathcal C_-$  
%\begin{widetext}
\[
\langle \mathcal C_\pm\,f_H|g_F\rangle =\langle f_H|\mathcal C_\mp\,g_F\rangle\,,\quad\qquad \forall f_H\in\Phi_H,\;\; \forall  g_F\in\Phi_F\,.
\]
%\end{widetext}
If  $g_F\in\Phi_F^\times$, the dual space of $\Phi_{F}$, the duality formula $\langle \mathcal C_\pm\,f_H|g_F\rangle = \langle f_H|\mathcal C_\pm^\dagger \,g_F\rangle$, defines a weakly continuous mapping $\mathcal C_\pm^\dagger$ from $\Phi_F^\times$  to $\Phi_H^\times$. This mapping obviously extends $\mathcal C_\mp$. The same is true if we interchange the role between the spaces. Resuming, we have
  \begin{eqnarray*}\label{5.15}
\mathcal C_\pm^\dagger : \Phi_H^\times \longmapsto \Phi_F^\times \qquad {\rm extend} \qquad \mathcal C_\mp :\Phi_F\longmapsto \Phi_H\,, \nonumber\\[2ex] 
\mathcal C_\pm^\dagger : \Phi_F^\times\longmapsto \Phi_H^\times \qquad {\rm extend} \qquad \mathcal C_\mp :\Phi_H\longmapsto \Phi_F\,.
\end{eqnarray*}
Similar relations hold for the other pairs $\mathcal D_\pm,\mathcal E_\pm$ and $\mathcal F_\pm$.

These results obtained for the generators of the Lie algebra $su(2,2)$ can be extended to its Universal Enveloping Algebra, UEA[$su(2,2)$], in the sense that all elements of UEA[$su(2,2)$] are continuous on $\Phi_{H\oplus F}$
 and can be continuously extended to weakly continuous mappings on $\left(\Phi_{H\oplus F}\right)^\times$. 
 
 As the representation of $su(2,2)$ is irreducible the UEA  is the space of continuous operators acting on the  RHS 
 $\Phi_{H\oplus F}\subset \mathcal H_{H\oplus F} \subset  \Phi_{H\oplus F}^\times$. Note that the UEA has a basis made of the monomials 
%\begin{widetext}
\begin{equation}\label{5.t}
 \mathcal T=\mathcal A_+^{\alpha}\,\mathcal B_+^{\beta}\,\mathcal C_+^{\gamma}\,\mathcal D_+^{\delta}\,
 \mathcal E_+^{\varepsilon}\,\mathcal F_+^{\zeta}\,\mathcal J^{\eta}\,\mathcal M^{\theta}\,\mathcal Q^{\iota}\,\mathcal F_-^{\zeta '}\,\mathcal E_-^{\varepsilon'}\,\mathcal D_-^{\delta'}\,\mathcal C_-^{\gamma'}\,\mathcal B_-^{\beta'}\,\mathcal A_-^{\alpha'}\,,
\end{equation}%\end{widetext}
and
it can be splitted in two subsets $UEA=UEA_1\oplus UEA_2$,  where $UEA_1$ and  $UEA_2$ are spanned by the operators $ \mathcal T_1$ and $ \mathcal T_2$, respectively,  such that 
\[
\gamma+\delta+\varepsilon+\zeta+\zeta '+\varepsilon'+\delta'+\gamma'\,=\,\left\{
\begin{array}{l}
\text{even} \quad \text{for}\;  \mathcal T_1\\[0.4cm]
\text{odd} \quad \text{for}\;  \mathcal T_2
\end{array}\right.\,.
\]
Thus, $UEA_1$ (that is an algebra)  is the space of continuous operators acting on the RHS
$\Phi_{H}\subset \mathcal L_{H} \subset \Phi_{H}^\times$ and 
$\Phi_{F}\subset \mathcal L_{F} \subset   \Phi_{F} ^\times$,
 i.e, in the space of the Jacobi Harmonics and  the Jacobi Fermion Harmonics, respectively.
Indeed 
%\begin{widetext}
\[\begin{array}{l}
 \mathcal T_1\,\Phi_H\subset  \Phi_H\,,\; \mathcal T_1^\dagger\,\Phi_H^\times\subset  \Phi_H^\times\,;\qquad\qquad 
 \mathcal T_1\,\Phi_F\subset  \Phi_F\,,\;
 \mathcal T_1^\dagger\,\Phi_F^\times\subset  \Phi_F^\times\,.
\end{array}\]%\end{widetext}
 However, $UEA_2$ mix the two RHS.

%%%%%%%%%%%%%%%%%%%%%%%%%%%
%%%%%%%%%%%%%%%%%%%%%%%%%%%

\subsection{Discrete and discrete-continuous bases in $\mathcal L_H$ }

In Section~\ref{discretecontinuousbases}, we have defined the continuous basis $|x,(m,q)\rangle$ for $m$ and $q$  fixed. Now, it is the time for extending these notions to changing values of $m$ and $q$.
% Integer values of $(j,m,q)$ are related to the Jacobi Harmonics and half-integers to  the Jacobi Fermion Harmonics, that allow to construct two separated but equivalent formalisms. 
 In this Subsection we will discuss only the case of  integer values of $(j,m,q)$ related to the Jacobi Harmonics.
 %, so that the index ``H'' is always assumed but it not will appear explicitely. 
 The analysis of the case with half-integer parameters is completely similar and it will be omitted.

Let   $\Xi_H$  be  the space of functions $f(x,\phi,\chi)\in\mathcal L_H$,
\begin{equation}\label{5.18}
f(x,\phi,\chi)=\sum_{j=0}^\infty\;\sum_{m=-j}^{j}\;\sum_{q=-j}^{j} f^j_{m,q} \, \mathcal N_j^{m,q}(x,\phi,\chi)\,,
\end{equation}
such that for any $r,s=0,1,2,\dots$, the following series converge:
\begin{equation}\label{5.19}
t_{r,s}(f):=\sum_{j,m,q} |f^j_{m,q}| \, (j+|m|+1)^r\,(j+|q|+1)^s\,.
\end{equation}
Due to \eqref{54} and \eqref{5.19}, the convergence of the series \eqref{5.18} is absolute and uniform due to the Wiesstrass M-Theorem, and hence pointwise. Using arguments  similar  to those in Section~\ref{discretecontinuousbases} we conclude that $\Xi_H\subset\mathcal L_H\subset \Xi_H^\times$ is a RHS. We also have that $\Xi_H\subset \Phi_H \subset \mathcal L_H \subset \Phi_H^\times \subset \Xi_H^\times$, where the canonical inclusions $\mathfrak i(f)=f$ are always continuous.

Next, let us consider an abstract infinite dimensional separable Hilbert space $\mathcal H_H$ and a unitary mapping $U:\mathcal H_H \longmapsto \mathcal L_H$. Take,  $\Omega_H\equiv U^{-1}\,\Xi_H$ and $\Psi_H\equiv U^{-1}\,\Phi_H$ with the topology transported by $U^{-1}$. Then, we have an abstract   RHS $ \O_H\subset \mathcal H_H \subset \O_H^\times$ related to
$\Xi_H\subset\mathcal L_H\subset \Xi_H^\times$ through  the following
the  diagram
\begin{equation}\label{5.diagram}
{U}\hskip-0.35cm\begin{array}{lllccccclll}
&\Omega_H&\subset &\Psi_H &\subset &\mathcal H_H &\subset & \Psi_H^\times&\subset & \Omega_H^\times
\\[0.2cm]
 & \downarrow & &\hskip-0.4cm U\downarrow && \hskip-0.4cm {U}\downarrow &&\hskip-0.445cm {U}\downarrow && \hskip-0.35cm {U}\downarrow
\\[0.2cm]
&\Xi _H&\subset &\Phi_H &\subset &{\mathcal L}_H&\subset \ & \Phi_H^\times&\subset &\Xi_H^\times
\end{array}\,.
\end{equation}
%where  $\Omega^\times \equiv U^{-1}\,\Xi^\times$. and $\Psi^\times \equiv U^{-1}\,\Phi^\times$.

Now, we generalize the kets $|j,(m,q)\rangle$   \eqref{jmq} %for any $j$ and any $m,q=-j,-j+1,\dots,j-1,j$ 
as follows
\begin{equation}\label{5.20}
|j,m,q\rangle:= U^{-1} \,\mathcal N^{m,q}_j(x,\phi,\chi)\,.
\end{equation}
From \eqref{5.2} the orthogonality and completeness relations \eqref{44} are now
\begin{equation}\label{5.44}
\langle j,m,q|j',m',q'\rangle =\delta_{j,j'}\,\delta_{m,m'}\,\delta_{q,q'}\,,\qquad \sum_{j,m,q} |j,m,q\rangle\langle j,m,q| = \mathcal I\,,
\end{equation}
where $\mathcal I$ is the identity on the space of the Jacobi Harmonics, $\mathcal L_H$.

%This definition is related to the previous one for $m,q$ fixed  \eqref{jmq}. 
Also for the elements $|x,(m,q)\rangle$ (with $x\in\mathbb [-1,1]$)  given by \eqref{45}, we can remove the parenthesis and define %$|x,m,q\rangle$, for any $ m$ and $q$  (integer) using  \eqref{45} and \eqref{5.20} obtaining that
\begin{equation}\label{5.45}
|x,m,q\rangle := \sum_j |j,m,q\rangle \,
\mathbb J_j^{m,q}(x)\,.
\end{equation}

From \eqref{5.44} and   \eqref{5.45} we obtain the relation
%In particular, definition \eqref{55} yields:
\begin{equation}\label{5.22}
\langle j,m',q'|x,m,q\rangle = \langle x,m,q|j,m',q'\rangle =
\mathbb J_j^{m,q}(x)\,\delta_{m,m'}\,\delta_{q,q'} \,,
\end{equation}
that shows that the $J_j^{m,q}(x)$ are the (real)  transition matrices between both bases.

The action of $|x,m,q\rangle$ on an arbitrary $|f\rangle\in\ \Omega_H$ is also given by \eqref{55} but  now   considering besides the sum in $m$ and $q$. After \eqref{56}, we show that $|x,m,q\rangle$ is anti-linear. Furthermore, it is also continuous on $\Omega_H$, since,
%\begin{widetext}
\begin{equation*}\begin{array}{lll}\label{5.21}
|\langle f|x,m,q\rangle|  &\le & \displaystyle \sum_{j,m,q} \big|f_{m,q}^j\big| \, (j+|m|+1)^2(j+|q|+1)\\[3ex]
&\le & \displaystyle \sum_{j,m,q} \big|f_{m,q}^j\big| \, (j+|m|+1)^2(j+|q|+1) =t_{2,1}(f)\,.
\end{array}\end{equation*}%\end{widetext}

Again from \eqref{0010}, \eqref{5.44} and \eqref{5.45} we easily see that
\begin{equation}\begin{array}{lll}\label{5.23}
\langle x',m',q'|x,m,q\rangle %&=&\displaystyle \sum_{j,m,q} \langle x',m',q'|j,m,q\rangle \,
%\mathbb J_j^{m,q}(x) \\[3ex]
%& =&\displaystyle \sum_{j}  
%\mathbb J_j^{m',q'}(x')\, \mathbb J_j^{m,q}(x)\,\delta_{m,m'}\,\delta_{q,q'} \\[2ex]  
%& =& \ds \sum_j \mathcal N^{m',q}_j (x',\phi,\chi) \, \mathcal N^{m,q}_j(x,\phi,\chi) \, \delta_{m,m'}\,\delta_{q,q'}  \\[2ex]  
& =& \delta(x-x')\,\,\delta_{m,m'}\,\delta_{q,q'}\,,
\end{array}\end{equation}

%We start with multiplying  \eqref{45} to the left by $\langle x',m',q'|$. Then,  we have
%%\begin{widetext}
%\begin{equation}\begin{array}{lll}\label{5.23}
%\langle x',m',q'|x,m,q\rangle &=&\displaystyle \sum_{j,m,q} \langle x',m',q'|j,m,q\rangle \,
%\mathbb J_j^{m,q}(x) \\[3ex]
%& =&\displaystyle \sum_{j}  
%\mathbb J_j^{m',q'}(x')\, \mathbb J_j^{m,q}(x)\,\delta_{m,m'}\,\delta_{q,q'} \\[2ex]  
%& =& \ds \sum_j \mathcal N^{m',q}_j (x',\phi,\chi) \, \mathcal N^{m,q}_j(x,\phi,\chi) \, \delta_{m,m'}\,\delta_{q,q'}  \\[2ex]  
%& =& \delta(x-x')\,\,\delta_{m,m'}\,\delta_{q,q'}\,,
%\end{array}\end{equation}
%where the identity between the first column and the second is due to \eqref{5.22} and between the second one the last term in \eqref{5.23} is just a consequence of \eqref{10} and \eqref{NAJF}.
%
On the other hand note that we can rewrite \eqref{5.22} using \eqref{5.23} as
 %$\langle j,m',q'|x,m,q\rangle=\langle x,m,q|j,m',q'\rangle$, since the AJF are real. 
\begin{equation*}\begin{array}{lll}\label{5.24}
\langle x,m,q|j,m',q'\rangle & =&\displaystyle \int_{-1}^1 \delta(x-x') \,\delta_{m,m'}\,\delta_{q,q'}\,
\mathbb J_j^{m,q}(x')  \,dx'\\[2ex]
&=&\displaystyle \int_{-1}^1 \langle x,m,q|x',m',q'\rangle\, 
\mathbb J_j^{m,q}(x')  \,dx'\,,
\end{array}\end{equation*}
and if we omit the arbitrary ket, $\langle x,m,q|$, we recover the generalization of \eqref{47}, i.e.
\begin{equation}\label{5.240}
|j,m,q\rangle =\int_{-1}^1 |x,m,q\rangle\, 
\mathbb J_j^{m,q}(x)  \,dx\,,
\end{equation}

Next, take an arbitrary $f\in\Psi_H$ and with the help of \eqref{5.240} we get
\begin{equation}\begin{array}{lll}\label{5.25}
\langle j',m',q'|f\rangle &=&\displaystyle \sum_{j,m,q} f^j_{m,q} \langle j',m',q'|
j,m,q\rangle \\[2ex]
&=&\ds  \sum_{j,m,q} f^j_{m,q} \int_{-1}^1 \langle j',m',q'| x,m,q\rangle \, \mathbb J_j^{m,q}(x)\,dx\\[2ex]
%&=&\ds \sum_{j,m,q} f^{j}_{m,q} \int_{-1}^1 \mathbb J_{j'}^{m',q'}(x) \,\mathbb J_j^{m,q}(x)\,dx\,.
 &=&\ds  \sum_{m,q} \int_{-1}^1 \langle j',m',q'| x,m,q\rangle \sum_j f^j_{m,q}\, 
\mathbb J_j^{m,q}(x)\,dx \\[2ex]
&=&\ds \sum_{m,q} \int_{-1}^1 \langle j',m',q'| x,m,q\rangle \,f^{m,q}(x)\,dx\,,
\end{array}\end{equation}
where the expression for $f^{m,q}(x)$ is obvious after \eqref{55}. We have used the Lebesgue theorem in order to insert the sum inside the integral from   \eqref{54} and \eqref{5.19}, Same expressions guarantee the absolute convergence of the resulting series, so that we may change the order of summation. 
%In consequence, equation \eqref{5.25} yields,
%\begin{equation*}\begin{array}{lll}\label{5.26}
%\langle j',m',q'|f\rangle &=&\displaystyle  \sum_{m,q} \int_{-1}^1 \langle j',m',q'| x,m,q\rangle \sum_j f^j_{m,q}\, 
%\mathbb J_j^{m,q}(x)\,dx \\[2ex]
%&=&\displaystyle \sum_{m,q} \int_{-1}^1 \langle j',m',q'| x,m,q\rangle \,f^{m,q}(x)\,dx\,,
%\end{array}\end{equation*}
 If we omit the arbitrary bra $\langle j',m',q'|$, we obtain the following  expression:
\begin{equation}\label{5.27}
|f\rangle = \sum_{m,q} \int_{-1}^1 dx\,|x,m,q\rangle\,f^{m,q}(x)\,,
\end{equation}
which generalizes \eqref{49}. From \eqref{5.27} and \eqref{55}, we have that for $|f\rangle\in\Omega_H$,
\begin{equation*}\label{5.28}
%f \equiv 
|f\rangle = \sum_{m,q} \int_{-1}^1 dx\,|x,m,q\rangle\,f^{m,q}(x) = \sum_{m,q} \int_{-1}^1 dx\,|x,m,q\rangle \langle x,m,q|f\rangle\,,
\end{equation*}
and we obtain the completeness identity 
\begin{equation*}\label{5.29}
\sum_{m,q} \int_{-1}^1 dx\,|x,m,q\rangle \langle x,m,q| = \mathcal I\,,
\end{equation*}
where $\mathcal I$ is the canonical injection $\mathcal I: \Omega_H\longmapsto \Omega_H^\times$.
Note that some of the proofs of the previous results are albeit formal.  

%Next, we intend to prove  that 
%formulas \eqref{46}-\eqref{50}. Some of these results generalize those formulas to fit with the  situation described in this subsection where    $m$ and $q$ are not fixed. 

Finally, the operators $\mathcal J,\mathcal M,\mathcal Q, \mathcal A_\pm,\mathcal B_\pm$ and $\mathcal T_1$  defined in \eqref{5.t} are continuous on $\Psi_H$ with formal adjoints having the same properties. In consequence, they can be extended into $\Psi_H^\times$ as weakly continuous operators. In addition, with the aid of the relation \eqref{53}, we show that these operators are continuous linear mappings from $\Omega_H$ into $\Phi_H$. The proof is simple. For instance for  $\mathcal A_+$, we readily see that $p_{r,s}(\mathcal A_+\,f)\le 2^{r/2}\,t_{r+1,s}(f)$, for all $f\in\Omega_H$, proving this continuity.

Let us recall again that the discussion for half-integer parameters 
 %and   Jacobi Fermion Harmonics 
is analogous.
%\end{widetext}
%%%%%%%%%%%%%%%%%%%
%%%%%%%%%%%%%%%%%%%
\subsection{The full continuous basis $\{|x,\phi,\chi\rangle\}$}\label{continuousbasis}

The next goal is to define  and give some properties of the generalized eigenvectors of the form $|x,\phi,\chi\rangle$. By definition $[\mathcal N_j^{m,q}(x,\phi,\chi)$ is the transition matrix between the bases $\{|j,m,q\rangle\}$ and 
$\{|x,\phi,\chi\rangle\}$, i.e., 
 for arbitrary  $|j,m,q\rangle\in\Xi_H$ we have
\begin{equation}\label{5C.1}
\langle x,\phi, \chi \big| j,m,q\rangle := \mathcal N_j^{m,q}(x,\phi,\chi)\,.
\end{equation}
If $|f\rangle =\ds\sum_{j=0}^\infty\sum_{m,q} f^j_{m,q} \,|j,m,q\rangle \in \Xi_H$, one has
\begin{equation}\begin{array}{lll}\label{5C.2}
|\langle f|x,\phi, \chi\rangle| & \le &\ds \sum_{j,m,q} |f^j_{m,q}|\, |\mathcal N_j^{m,q}(x,\phi,\chi)| \le \sum_{j,m,q} |f^j_{m,q}|\, \sqrt{j+1/2}\,|\mathcal J_j^{m,q}(x)| \\[2ex] 
&\le &\ds \sum_{j,m,q} |f^j_{m,q}|\,\sqrt{j+1/2}\,(j+|m|+1)(j+|q|+1) \\[2ex] 
&\le &\ds\sum_{j,m,q} |f^j_{m,q}|\,(j+|m|+1)^2(j+|q|+1)  = t_{2,1}(|f\rangle)\,.
\end{array}\end{equation}
The anti-linearity of $|x,\phi,\chi\rangle$ on $\Omega_H$ is obvious and the latter chain of inequalities shows that $|x,\phi,\chi\rangle$ is also continuous on $\Omega_H$, so that $|x,\phi,\chi\rangle \in \Omega_H^\times$. 

In Appendix~B \eqref{104} we have proven that $e^{-i\phi}(\cos\chi+iX \,\sin\chi)$ is a continuous mapping in $\Phi_{H,F}$ for each value of $\phi$ and $\chi$. 

%{\bf Vamos a usar provisionalmente la notaci\'on $\Xi_{H,F}$}. 
On each function $f(x,\phi,\chi)$ in the Hilbert spaces $\mathcal \mathcal L_H$, one may define the following bounded operators:
\begin{equation}\begin{array}{rllrll}\label{5C.3}
X\,f_{H,F}(x,\phi,\chi) &=& x\,f_{H,F}(x,\phi,\chi)\,,&\qquad \boldsymbol\Phi\,f_{H,F}(x,\phi,\chi)&=&\phi\,f_{H,F}(x,\phi,\chi)\,,\\[2ex] \cos{\boldsymbol\chi}\,f_{H,F}(x,\phi,\chi) &=&\cos \chi \,f_{H,F}(x,\phi,\chi)\,,
& \qquad \sin{\boldsymbol\chi}\,f_{H,F}(x,\phi,\chi)&=& \sin\chi\,f_{H,F}(x,\phi,\chi)\,.
\end{array}\end{equation}
Then, following the same lines as in Appendix~B, one finds that  the operator
\begin{equation}\label{5C.4}
e^{-i\boldsymbol\Phi} (\cos\boldsymbol\chi+iX\,\sin\boldsymbol\chi)
\end{equation}
is continuous on $\Xi_H$. Therefore, the formal adjoint is continuous on the weak anti-dual, $\Xi_H^\times$. 
Consequently, by means of  the operator $U$, defined in \eqref{5.diagram} we get that 
\begin{equation}\label{5C.5}
R= U^{-1}\,e^{-i\boldsymbol\Phi} (\cos\boldsymbol\chi+iX\,\sin\boldsymbol\chi)\, U\,,
\end{equation}
is a continuous operator on $\Omega_{H}$ and its formal adjoint is weakly continuous on $\Omega_H^\times$, so that
\begin{equation}\label{5C.6}
R^\dagger\, |x,\phi,\chi\rangle = e^{i\phi} (\cos\chi -ix\,\sin\chi)\,|x,\phi,\chi\rangle\,,
\end{equation}
for each $x\in[-1,1]$ and each value of the angles $\phi$ and $\chi$. 

%\subsection{Some formal expressions}

From \eqref{5C.1}, \eqref{5.2} and the completeness relation of \eqref{5.44} we easily obtain 
\begin{equation}\label{5C.7}
\langle x',\phi',\chi'|x,\phi,\chi\rangle =\delta(x-x')\,\delta(\phi-\phi')\,\delta(\chi-\chi')\,,
\end{equation}
and we immediately have that 
\begin{equation}\label{5C.8}
|j,m,q\rangle = \int \mathcal N_j^{m,q}(x,\phi,\chi)\,|x,\phi,\chi\rangle\,dx\,d\phi\,d\chi\,.
\end{equation}
If we sustitute $\mathcal N(x,\phi,\chi)$ for its expression  of \eqref{5C.1} we get
%From \eqref{5C.8} The unitarity of $U$ \eqref{5.diagram} shows that
\begin{equation}\label{5C.9}
|j,m,q\rangle = \ds \int |x,\phi,\chi\rangle\,\langle x,\phi, \chi \big|  j,m,q\rangle\,dx\,d\phi\,d\chi 
\end{equation}
%\begin{equation}\begin{array}{lll}\label{5C.9}
%|j,m,q\rangle &=& \ds \int |x,\phi,\chi\rangle\,\langle x,\phi, \chi \big|  j,m,q\rangle\,dx\,d\phi\,d\chi 
%\\[3ex] 
%&=&\ds \int |x,\phi,\chi\rangle \langle x,\phi,\chi |j,m,q\rangle \, dx\,d\phi\,d\chi \,,
%\end{array}\end{equation}
%
%From \eqref{5C.8} The unitarity of $U$ \eqref{5.diagram} shows that
%\begin{equation}\begin{array}{lll}\label{5C.9}
%\langle j',m',q'|j,m,q\rangle &=& \ds \int \mathcal [N^{m',q'}_{j'} (x,\phi,\chi)]^*\,\mathcal N^{m,q}_j(x,\phi,\chi)\,dx\,d\phi\,d\chi 
%\\[3ex] 
%&=&\ds \int \langle j',m',q'|x,\phi,\chi\rangle \langle x,\phi,\chi |j,m,q\rangle \, dx\,d\phi\,d\chi \,,
%\end{array}\end{equation}
so that
\begin{equation}\label{5C.10}
\int |x,\phi,\chi\rangle \langle x,\phi,\chi |  \, dx\,d\phi\,d\chi =\mathcal I\,,
\end{equation}
which is the canonical injection $\mathcal I_{}: \Omega_{H} \longmapsto \Omega^\times_{H}$. 
Next, recalling the completeness relation \eqref{5.44}
%\begin{equation}\label{5C.11}
%\sum_{j,m,q} |j,m,q\rangle\langle j,m,q| = I\,,
%\end{equation}
where $I$ is the identity on either $\mathcal L_H$, we have formally  for each value of $x$, $\phi$ and $\chi$ that
\begin{equation}\label{5C.12}
|x,\phi,\chi\rangle = \sum_{j,m,q} |j,m,q\rangle\langle j,m,q |x,\phi,\chi\rangle = \sum_{j,m,q} [\mathcal N_j^{m,q}(x,\phi,\chi)]^*\, |j,m,q\rangle\,.
\end{equation}
\end{widetext}
%%%%%%%%%%%%%%%%%%%%%%%%%%%%%%%%%%%%%%%%%%%%%%%%%%%
%%%%%%%%%%%%%%%%%%%%%%%%%%%%%%%%%%%%%%%%%%%%%%%%%%%

\section{Concluding remarks}\label{conclusions}

The set of  Algebraic Jacobi Functions, $\{\mathcal J_j^{m,q}(x)\}$ as well as the 
set of Normalized Algebraic Jacobi Functions $\{\mathbb J_j^{m,q}(x)\}$,  support a non-unitary and non-irreducible representation for the Lie algebra $su(2,2)$. 
However,  $\{\mathbb J_j^{m,q}(x)\}^{m,q \text{(fixed)}}_{j\ge \sup(|m|,|q|)}$ constitutes an orthonormal basis for the Hilbert space $L^2[-1,1]$ and supports a UIR of $su(1,1)$. These bases span dense subspaces of  $L^2[-1,1]$ having their own locally convex topology, stronger than the Hilbert space topology. Each of the subspaces is characterized by fixed values of $m$ and $q$, so that we may denote them as $\Phi_{m,q}$. The triplets $\Phi_{m,q}\subset L^2[-1,1] \subset \Phi_{m,q}^\times$ form RHS or Gelfand triplets, $\Phi_{m,q}^\times$ being the dual of $\Phi_{m,q}$. Each of the spaces $\Phi_{m,q}$ support a representation of the Lie algebra
$su(1,1)$ by continuous operators on $\Phi_{m,q}$, where the continuity is respect to the locally convex topology of these spaces. These operators may be extended by duality to continuous operators on the duals $\Phi_{m,q}^\times$, so that we have  different and non-equivalent representations of the algebra $su(1,1) $ by continuous operators.

Algebraic Jacobi Functions are also related to different pairs of creation and annihilation ladder operators transforming  AJF into others with different values of the parameters $m$ and $q$. Consequently, these ladder operators should be transformations between different spaces of the family $\Phi_{m,q}$. This is really what happens, the ladder operators being continuous as mappings between these spaces. These continuity induce similar relations between the duals.

Up to this point, we have used a concrete realization of RHS using Algebraic Jacobi Functions. For each of these concrete realizations, we may construct an abstract RHS, $\Psi_{m,q}\subset\mathcal H\subset \Psi_{m,q}^\times$ unitarily equivalent to the original $\Phi_{m,q}\subset L^2[-1,1]\subset \Phi_{m,q}^\times$. This abstract representation permits to span each of the function on the space $\Phi_{m,q}$, for whatever values of $m$ and $q$, in terms of continuous basis of vectors on the dual, $\Psi_{m,q}^\times$, of the abstract space, $\Psi_{m,q}$, unitarily equivalent to $\Phi_{m,q}$. These continuous bases are eigenvectors of some continuous operators defined on $\Psi_{m,q}$.

Along this continuous basis, we always have a discrete orthonormal basis in terms of which all vectors on the spaces  
$\Phi_{m,q}$ and $\Psi_{m,q}$ may be written. Since these spaces are contained in their respective duals, these orthonormal basis are also contained in the duals, so that both discrete and continuous basis coexist in the same dual spaces. Then, we have established the relation between discrete and continuous basis in our example and have shown that brackets among vectors of these two different basis are well defined and can be given in terms of the AJF.

Remark that we have used two non-equivalent choices for the locally convex topology of the spaces spanned by the AJF with fixed values of $m$ and $q$ and therefore, we have two non-equivalent triplets. These spaces, $\Phi_{m,q}$ and  $\Xi_{m,q}$, have not only different topologies, they are even different as linear spaces of functions, although one contains the other as subspace ($\Xi_{m,q}\subset \Phi_{m,q}$). Furthermore, some particular properties are different in each case. These spaces are defined by series of functions that always converge pointwise in one case ($\Xi_{m,q} $) and not in the other. Nevertheless, the properties of the operators representing the algebras $su(2,2)$ and $su(1,1)$, such as continuity as well as the properties of the ladder operators are similar in both triplets.

So far, we have obtained RHS for fixed  values of $m$ and $q$. However by introducing Jacobi Harmonic we have been able to construct a RHS   that include all values of the parameters. This construction allows also to split this RHS  into two different triplets, one for  integer and the other for  half-integer values of $j$. On these spaces, we may define all the objects treated along the present article and continuity properties may be also proved. In this context, we have discussed the behaviour and properties of ladder operators and continuous and discrete bases, relations between both triplets, etc.,  pointing out the relevant role played by  the   Lie algebra  and its  UEA.

A possible physical applications of AJF could be   to describe quantum systems of two  particles with the same spin because for  $j$ fixed they are described by the representation  $D_j\otimes D_j$  of $SU(2)\otimes  SU(2)$ (see \eqref{11}).

%%%%%%%%%%%%%%%%%%%%%%%%%%%%%%%%%%%%%%%%%%%
%%%%%%%%%%%%%%%%%%%%%%%%%%%%%%%%%%%%%%%%%%%%%

\begin{widetext}
\section*{Acknowledgements}

We acknowledge partial financial support to the Spanish MINECO, grant MTM2014-57129-C2-1-P, and the Junta de Castilla y Le\'on, grants  VA137G18 and BU229P18.

%%%%%%%%%%%%%%%%%%%%%%%%%%%%%%%%%%%%%%%%%%%%
%%%%%%%%%%%%%%%%%%%%%%%%%%%%%%%%%%%%%%%%%%%%
\appendix
\section*{Appendix A: Properties of the Algebraic Jacobi Functions.}\label{symmetries}
The AJF  \eqref{3} are either pure polynomials or ``quasi-polynomials'' in $x$. 
Indeed
\begin{equation*}\label{a00}\begin{array}{ll}
j,m,q\in \Z :\qquad
\qquad & \mathcal J^{m,q}_{j}(x)= \left\{ \begin{array}{rl}
P(x)\quad &\text{ if}\;\; m-q \;\;\text{even}
\\[0.4cm]
\sqrt{1-x^2}\, P(x)\quad &\text{ if}\;\; m-q \;\;\text{odd}
\end{array}
\right.
\\[0.75cm]
j,m,q\in \Z+1/2 :\qquad\qquad & \mathcal J^{m,q}_{j}(x)= \left\{ \begin{array}{rl}
\sqrt{1-x}\,P(x)\quad &\text{ if}\;\; m-q \;\;\text{even}
\\[0.4cm]
\sqrt{1+x}\, P(x)\quad &\text{ if}\;\; m-q \;\;\text{odd}
\end{array}
\right.
\end{array}
\end{equation*}
where $P(x)$ are  appropriate polynomials in $x$.

We mention in Section~\ref{jacobifunctions} that the AJF $\mathcal J^{m,q}_j(x)$ have  symmetry properties in   $(m,q,x)$ related to the Jacobi differential equation \eqref{jacobiequation} 
which  is invariant under    $(m,q,x)\leftrightarrow (q,m,x)$, 
$(m,q,x)\leftrightarrow (-m,-q,x)$, $(m,q,x)\leftrightarrow (m,-q,-x)$ and their combinations.

We can obtain the following explicit expression for the $\mathcal J^{m,q}_j(x)$ \eqref{5}
\begin{equation}\label{a1}
\mathcal J^{m,q}_j(x)= K[j,m,q]\, \ds \sum_{s=0}^{j-m}\frac{(-1)^{j-m-s}\,
 \left(\frac{1-x}{2} \right)^{j+\frac{q-m}{2}-s} \left(\frac{1+x}{2} \right)^{\frac{m-q}{2}+s}}{\Gamma(s+1)\,\Gamma(j-m-s+1)\,\Gamma(j+q-s+1)\,\Gamma(m-q+s+1)}
 \,,
\end{equation}
such that
\begin{equation*}\label{a2}
K[j,m,q]:=
\sqrt{{\Gamma(j+m+1)\,\Gamma(j-m+1)}\,{\Gamma(j+q+1) \,\Gamma(j-q+1)}}\,.
\end{equation*}
The expression of $K[j,m,q]$  
is invariant under the above mentioned changes interchanges of  $m$ and $q$.

From \eqref{a1} $\mathcal J^{q,m}_j(x)$ reads  as follows
\begin{equation}\label{a3}
\mathcal J^{q,m}_j(x)= K[j,m,q]\, \ds \sum_{s=0}^{j-q}\frac{(-1)^{j-q-s}\,
 \left(\frac{1-x}{2} \right)^{j+\frac{m-q}{2}-s} \left(\frac{1+x}{2} \right)^{\frac{q-m}{2}+s}}{\Gamma(s+1)\,\Gamma(j-q-s+1)\,\Gamma(j+m-s+1)\,\Gamma(q-m+s+1)} \,.
\end{equation}
Comparing with  expression \eqref{a1} of  $\mathcal J^{m,q}_j(x)$ we see that with the change $s=q-m+t$ we transform
\eqref{a1} in
\begin{equation}\label{a4}
\mathcal J^{m,q}_j(x)= K[j,m,q]\, \ds \sum_{t=m-q}^{j-q}\frac{(-1)^{j-q-t}\,
 \left(\frac{1-x}{2} \right)^{j+\frac{m-q}{2}-t} \left(\frac{1+x}{2} \right)^{\frac{q-m}{2}+t}}{\Gamma(q-m+t+1)\,
 \Gamma(j-q-t+1)\,\Gamma(j+m-t+1)\,\Gamma(t+1)} \,,
\end{equation}%\end{widetext}
Now let us consider the possible values of $m-q$. If $m-q\geq 0$ (i.e., $m\geq q$) we observe that $q-m+t+1\leq 0$ for 
$t=0,1,\cdots, m-q-1$ then $|\Gamma(q-m+t+1)|=\infty$ and these terms in \eqref{a4} shall become zero. In the opposite case of $m-q\leq 0$ (i.e., $m\leq q$)  we see $t+1\leq 0$ for 
$t= m-q, m-q+1,\cdots, -1$ and hence  $|\Gamma(t+1)|=\infty$. So, in both cases we can consider the    sum for $t=0,1,\cdots, j-q$  recovering the expression  \eqref{a3} for $\mathcal J^{q,m}_j(x)$. Then we have proved that 
$\mathcal J^{m,q}_j(x)=\mathcal J^{q,m}_j(x)$.

Now we study the case of  $\mathcal J^{-m,-q}_j(x)$ which expression from \eqref{a1} is
%\begin{widetext}
\begin{equation*}\label{a5}
\mathcal J^{-m,-q}_j(x)= K[j,m,q]\, \ds \sum_{s=0}^{j+m}\frac{(-1)^{j+m-s}\,
 \left(\frac{1-x}{2} \right)^{j+\frac{-q+m}{2}-s} \left(\frac{1+x}{2} \right)^{\frac{-m+q}{2}+s}}{\Gamma(s+1)\,\Gamma(j+m-s+1)\,\Gamma(j-q-s+1)\,\Gamma(-m+q+s+1)}
\end{equation*}%\end{widetext}
Comparing this expression with that of $\mathcal J^{q,m}_j(x)$ \eqref{a3} we see that  both coincide up a global sign 
$(-1)^{q+m}$ and the upper limit of the sum that is $j+m$ instead  $j-q$. But considering, for instance, that $j+m\geq j-q$   we see that   $|\Gamma(j-q-s+1)|=\infty$  for 
$s= j-q+1, j-q+2,\cdots, j+m$.  In the opposite case  $j+m\leq j-q$ we have $|\Gamma(j+m-s+1)|=\infty$  for 
$s= j+m+1, j+m+2,\cdots, j-q$. So, in both cases we can consider the upper limit of the sum $s-j-q$ and we have proved 
$\mathcal J^{-m,-q}_j(x)=(-1)^{m+q}\,\mathcal J^{q,m}_j(x)$..

On the other hand, from the inspection of the Jacobi differential equation \eqref{jacobiequation} we see that under the change $m\to -m$ (or 
$q\to -q$, $m\leftrightarrow -q$) we recover the equation \eqref{jacobiequation}  if we also perform the change of sign of the variable $x$, i.e., $x\to -x$. For instance,   $\mathcal J^{m,-q}_j(-x)$ is solution of \eqref{jacobiequation} and
it will be written as
%\begin{widetext}
\begin{equation*}\label{a10}
\mathcal J^{m,-q}_j(-x):= K[j,m,q]\, \ds \sum_{s=0}^{j-m}\frac{(-1)^{j-m-s}\,
 \left(\frac{1+x}{2} \right)^{j+\frac{-q-m}{2}-s} \left(\frac{1-x}{2} \right)^{\frac{m+q}{2}+s}}{\Gamma(s+1)\,\Gamma(j-m-s+1)\,\Gamma(j-q-s+1)\,\Gamma(m+q+s+1)} \,.
\end{equation*}
Using the change $s=j-m-t$ we rewrite the previous expressions as
\begin{equation*}\label{a11}
\mathcal J^{m,-q}_j(-x)= K[j,m,q]\, \ds \sum_{t=0}^{j-m}\frac{(-1)^{t}\,
 \left(\frac{1+x}{2} \right)^{\frac{m-q}{2}+t} \left(\frac{1-x}{2} \right)^{j+\frac{q-m}{2}+t}}{\Gamma(j-m-t+1)\,\Gamma(t+1)\,\Gamma(m-q+t+1)\,\Gamma(j+q-t+1)}
 \,.
\end{equation*}%\end{widetext}
Since $(-1)^{j-m-t}=(-1)^{j-m}\,(-1)^{-t}=(-1)^{j-m}\,(-1)^{t}$ we see that 
$\mathcal J^{m,-q}_j(-x)=(-1)^{j-m}\,\mathcal J^{m,q}_j(x)$.

Resuming we have for all $m$ and $q$
\begin{equation}\label{a9}\begin{array}{l}
\mathcal J^{q,m}_j(x)=\mathcal J^{m,q}_j(x)
\\[0.4cm]
J^{-m,-q}_j(x)=(-1)^{m+q}\,\mathcal J^{m,q}_j(x) \,,
\\[0.4cm]
\mathcal J^{m,-q}_j(-x)=(-1)^{j-m}\,\mathcal J^{m,q}_j(x)\,,
\end{array}
\end{equation}
and their combinations.

%%%%%%%%%%%%%%%%%%%%%%%%%
%%%%%%%%%%%%%%%%%%%%%%%%%
\section*{Appendix B: Multiplication operators.}\label{someoperator}

In \cite{CGO1}, we have analyzed the behaviour of the multiplication operator on spaces including Spherical Harmonics. Because the relations between Spherical Harmonics and Jacobi Harmonics we  study now the action of the multiplication operator on spaces of Jacobi functions.  

Starting from the definition of the functions $\mathcal J_j^{m,q}(x)$ given in  \eqref{6},  and the property for the Jacobi polynomials given in \cite{abramovich-stum} (p. 782, eq. 22.7.15)  as
%\begin{widetext}
\begin{equation}\label{89}
(1-x)\,(n+\frac{\alpha}{2}+\frac{\beta}{2}+1)\, J^{\alpha +1,\beta}_{n}(x)=
(n+\alpha+1)\, J^{\alpha,\beta}_{n}(x)-(n+1)\, J^{\alpha,\beta}_{n+1}(x)\,,
\end{equation}
we obtain the following  properties of the multiplication operator $Xf(x)=xf(x)$ on the AJF  $\mathcal J_j^{m,q}(x)$:
\begin{equation}\label{90}
x\,\mathcal J^{m,q}_j(x)= \mathcal J^{m,q}_j(x)+c^{m,q}_j\,\mathcal J^{m-1,q-1}_{j-1}(x)
+d^{m,q}_j\,\mathcal J^{m-1,q-1}_{j}(x)+e^{m,q}_j\,\mathcal J^{m-1,q-1}_{j+1}(x)\,,
\end{equation}
with
\begin{equation}\label{91}
\begin{array}{l}
c^{m,q}_{j}=\displaystyle 
-\frac{\sqrt{(j+m-1)\,(j+m)\,(j+q-1)\,(j+q)}}{j\,(2j+1)}\,,\\[3ex]
d^{m,q}_{j}=\displaystyle 
\frac{\sqrt{(j-m+1)\,(j+m)\,(j-q+1)\,(j+q)}}{j\,(j+1)\,(2j+1)}\,,\\[3ex]
e^{m,q}_{j}=\displaystyle 
-\frac{\sqrt{(j-m+1)\,(j-m+2)\,(j-q+1)\,(j-q+2)}}{(j+1)\,(2j+1)}\,.
\end{array}\end{equation}
In the case of the functions $\mathbb J_j^{m,q}(x)$ defined in \eqref{NAJF} expression \eqref{90} becomes
\begin{equation}\label{900}
\begin{array}{l}
\ds x\,\mathbb J^{m,q}_j(x)= \mathbb J^{m,q}_j(x)+c^{m,q}_j\,\sqrt{\frac{j+1/2}{j-1/2}}\,\mathbb J^{m-1,q-1}_{j-1}(x)
\\[0.4cm]
\hskip3cm
\ds+\, d^{m,q}_j\,\mathbb J^{m-1,q-1}_{j}(x)+e^{m,q}_j\,\sqrt{\frac{j+1/2}{j+1/2}}\,\mathbb J^{m-1,q-1}_{j+1}(x)\,.
\end{array}\end{equation}

Now we have to look for   the spaces where the multiplication operator $X$ is well defined and continuous. Obviously, $X$ mixes up the spaces $\Xi_{m,q}$ and also spaces $\Phi_{m,q}$, so that we need another kind of space. The fact that  $X$ conserves the parity of $j,m$ and $q$ gives us an interesting clue for the definition of these spaces. 

Let us consider two Hilbert spaces, one for $j$ integer and the other for $j$ half-integer that we denote here as $\mathcal G_H$ and $\mathcal G_F$, respectively. Thus, $\mathcal G_H$ is the space of all series of the form:
\begin{equation}\label{92}
f(x):= \sum_{j=0}^\infty \sum_{m,q=-j}^j f_j^{m,q} \mathbb J_j^{m,q}(x)\,,\qquad {\rm with} \qquad  \sum_{j=0}^\infty \sum_{m,q=-j}^j \big|f_j^{m,q}\big|^2<\infty\,,
\end{equation}
and a similar definition for $j$ half-integer (in this case the sum goes from $j=1/2$ to $\infty$). These spaces may be written as infinite direct sums:
\begin{equation*}\label{93}
\mathcal G_H =\bigoplus_{|m|,|q|=0}^\infty \mathcal G_{|m|,|q|}\,, \qquad \mathcal G_F = \bigoplus_{|m|,|q|=1/2}^\infty \mathcal G_{|m|,|q|}\,,
\end{equation*}
where $\mathcal G_{|m|,|q|}$ are copies of $L^2[-1,1]$. Then, let us define the space $\Delta_H$ as the space of all series in $\mathcal G_H$ such that
\begin{equation}\label{94}
||f||^2_{r,s}:= \sum_{j=0}^\infty \sum_{m,q=-j}^j \big|f_j^{m,q}\big|^2 \,(j+|m|+1)^{2r}\,(j+|q|+1)^{2s}<\infty\,,
\end{equation}
for all $r,s=0,1,2,\dots$. Analogously, we define $\Delta_F$ for $j$ half-integer. 

Then, the multiplication operator $X$ is continuous on both $\Delta_H$ and $\Delta_F$. This means that both $\Delta_{H,F}$ reduce $X$, i.e., $X\Delta_H\subset \Delta_H$ and $X\Delta_F\subset \Delta_F$ and that $X$ is a continuous linear mapping on $\Delta_{H,F}$. Let us proceed for the proof for $\Delta_H$ (the proof for $\Delta_F$ being identical). Observe that the two first coefficients in \eqref{91}, $c_j^{m,q}$ and $d_j^{m,q}$, have a $j$ in the denominator, which may be a problem for $j=0$. This is not the case, since after \eqref{90} and \eqref{91}, we
have that  
\begin{equation*}\label{950}
c_0^{m,q}=d_0^{m,q}=0\,,
\end{equation*} 
or in other words 
\begin{equation*}\label{95}\begin{array}{lllll}
X\,\mathcal J_0^{0,0}(x)&=& x\,\mathcal J_0^{0,0}(x) &=& \mathcal J_0^{0,0}(x) -\sqrt 2\,\mathcal J_1^{-1,-1}(x)\,,
\\[0.4cm]
X\,\mathbb J_0^{0,0}(x)&=& x\,\mathbb J_0^{0,0}(x) &=& \mathbb J_0^{0,0}(x) -\frac{2\sqrt 2}{\sqrt{3}}\,\mathbb J_1^{-1,-1}(x)\,.
\end{array}\end{equation*}
So that, for $j=0$, the coefficients having a factor $j$ in the denominator do not appear. Then, for any series \eqref{92} in $\Delta_H$, we have that
\begin{equation*}\begin{array}{lll}\label{96}
(X\,f)(x)&=&\ds X\left[ \sum_{j=0}^\infty \sum_{m,q=-j}^j f_j^{m,q} \mathbb J_j^{m,q}(x) \right] = \sum_{j=0}^\infty \sum_{m,q=-j}^j f_j^{m,q} \,x\,\mathbb J_j^{m,q}(x) \\[3ex] 
&=&\ds \sum_{j,m,q}
f_j^{m,q} \mathbb J_j^{m,q}(x)  + \sum_{j,m,q}
\sqrt{\frac{j+1/2}{j-1/2}}\,f_j^{m,q} c_j^{m,q} \mathbb J_{j-1}^{m-1,q-1}(x)\\[3ex] 
&&\quad\ds+ \sum_{j,m,q} 
f_j^{m,q} \,d_j^{m,q}\, \mathbb J_j^{m-1,q-1} (x) + \sum_{j,m,q}
 \sqrt{\frac{j+1/2}{j+3/2}}\,f_j^{m,q}\, e_j^{m,q}\,\mathbb J_{j+1}^{m-1,q-1}(x)\,.
\end{array}\end{equation*}
Since $j+|m|+2 \le 2(j+|m|+1)$, all coefficients \eqref{91} are bounded by $2(j+|m|+1)(j+|q|+1)$ (recall above comments for the case $j=0$)  and taking into account \eqref{94} and  \eqref{540} 
we have that
\begin{equation*}\begin{array}{lll}\label{97}
||Xf||^2_{r,s} &\le & \ds 16 \sum_{j=0}^\infty \sum_{m,q=-j}^j \big|f_j^{m,q}\big|^2\,(j+1/2) 
\,(j+|m|+1)^{2(r+1)}\,(j+|q|+1)^{2(s+1)} 
\\[0.4cm]\ds&\le & \ds 
16 \sum_{j=0}^\infty \sum_{m,q=-j}^j \big|f_j^{m,q}\big|^2 
\,(j+|m|+1)^{2(r+2)}\,(j+|q|+1)^{2(s+1)} = 16 \, ||f||^2_{r+2,s+1}\,.
\end{array}\end{equation*}
This proves both, that $X$ preserves $\Delta_H$, $X\Delta_H\subset \Delta_H$ and that $X$ is continuous on $\Delta_H$. Same proof is valid for $\Delta_F$. 

Next, let us see what we may say in relation with the spaces spanned by the functions $\mathcal N_j^{m,q}(x,\phi,\chi)$ and the multiplication operator $X$. First of all, we have to remark that by the  symmetry property under the interchange 
$\alpha\;\leftrightarrow\; \beta$  that reads
\[
 J^{\beta,\alpha}_{n}(x)= (-1)^n\,J^{\alpha,\beta}_{n}(-x)\,.
 \]
Then eq. \eqref{89} may be also written as
\begin{equation}\label{98}
(1+x)\,(n+\frac{\alpha}{2}+\frac{\beta}{2}+1)\, J^{\alpha,\beta +1}_{n}(x)=
(n+\beta+1)\,  J^{\alpha,\beta}_{n}(x)+(n+1)\,  J^{\alpha,\beta}_{n+1}(x)\,.
\end{equation}
Note that  \eqref{98} appears also in Ref.÷\cite{abramovich-stum} (p. 782, eq. 22.7.16).
From \eqref{98}, we readily obtain the following recurrence relation for the AJF:
\begin{equation}\label{99}
x\,\mathcal J^{m,q}_j(x)=- \mathcal J^{m,q}_j(x)+\hat c^{m,q}_j\,\mathcal J^{m-1,q+1}_{j-1}(x)
+\hat d^{m,q}_j\,\mathcal J^{m-1,q+1}_{j}(x)+\hat e^{m,q}_j\,\mathcal J^{m-1,q+1}_{j+1}(x)\,,
\end{equation}
with
\begin{equation*}\label{100}
\begin{array}{l}
\hat c^{m,q}_{j}=\displaystyle 
\frac{\sqrt{(j+m-1)\,(j+m)\,(j-q-1)\,(j-q)}}{j\,(2j+1)}\,,\\[3ex]
\hat d^{m,q}_{j}=\displaystyle 
\frac{\sqrt{(j-m+1)\,(j+m)\,(j-q)\,(j+q+1)}}{j\,(j+1)\,(2j+1)}\,,\\[3ex]
\hat e^{m,q}_{j}=\displaystyle 
\frac{\sqrt{(j-m+1)\,(j-m+2)\,(j+q+1)\,(j+q+2)}}{(j+1)\,(2j+1)}\,.
\end{array}\end{equation*}
The corresponding version of the expression \eqref{99} for the NAJF $\mathbb J_j^{m,q}(x)$ is
\begin{equation}\label{901}
\begin{array}{l}
\ds x\,\mathbb J^{m,q}_j(x)= -\mathbb J^{m,q}_j(x)+\hat c^{m,q}_j\,\sqrt{\frac{j+1/2}{j-1/2}}\,\mathbb J^{m-1,q+1}_{j-1}(x)
\\[0.4cm]
\hskip3cm
\ds+\, \hat d^{m,q}_j\,\mathbb J^{m-1,q+1}_{j}(x)+\hat e^{m,q}_j\,\sqrt{\frac{j+1/2}{j+1/2}}\,\mathbb J^{m-1,q+1}_{j+1}(x)\,.
\end{array}\end{equation}

From \eqref{90} and \eqref{99}, we obtain that
\begin{equation*}\begin{array}{lll}\label{101}
X\,\mathcal J_j^{m,q}(x) = x\,\mathcal J^{m,q}_j(x)&=&\displaystyle
\frac{c^{m,q}_j}{2}\,\mathcal J^{m-1,q-1}_{j-1}(x)+\frac{\hat c^{m,q}_j}{2}\,\mathcal J^{m-1,q+1}_{j-1}(x)
+\frac{d^{m,q}_j}{2}\,\mathcal J^{m-1,q-1}_{j}(x) ) \\[2ex]
&&\hskip.25cm \displaystyle +\frac{\hat d^{m,q}_j}{2}\,\mathcal J^{m-1,q+1}_{j}(x)+\frac{e^{m,q}_j}{2}\,\mathcal J^{m-1,q-1}_{j+1}(x)+\frac{\hat e^{m,q}_j}{2}\,\mathcal J^{m-1,q+1}_{j+1}(x)\,.
\end{array}\end{equation*}

Then, let us use the definition of the functions $\mathcal N_j^{m,q}(x,\phi,\chi)$ as given in \eqref{5.1} and \eqref{900}, so as to obtain the following relation:
\begin{equation*}\begin{array}{l}\label{102}
e^{-i \phi}\,e^{-i \chi}\,(1-x)\,\mathcal N^{m,q}_j(x,\phi,\chi) \\[2ex] =\displaystyle
\sqrt{\frac{j+1/2}{j-1/2}}\,c_j^{m,q}\,  \mathcal N^{m-1,q-1}_{j-1}(x,\phi,\chi)  
 +d_j^{m,q}\,  \mathcal N^{m-1,q-1}_{j}(x,\phi,\chi)+\sqrt{\frac{j+1/2}{j+3/2}}\,e_j^{m,q}\,  \mathcal N^{m-1,q-1}_{j+1}(x,\phi,\chi)\,.
\end{array}\end{equation*}
Now, it is rather easy to show that the operator $e^{-i\phi}\,e^{-i\chi} (1-X)$ is a continuous linear operator on $\Phi_H$ and $\Phi_F$ as defined in Subsection~\ref{5.A}. Similarly, we may find another similar relation such as
\begin{equation*}\begin{array}{l}\label{103}
e^{-i \phi}\,e^{i \chi}\,(1+x)\,\mathcal N^{m,q}_j(x,\phi,\chi)\\[2ex]
\displaystyle
=\sqrt{\frac{j+1/2}{j-1/2}}\,\hat c_j^{m,q}\,  \mathcal N^{m-1,q+1}_{j-1}(x,\phi,\chi) 
 +\hat d_j^{m,q}\,  \mathcal N^{m-1,q+1}_{j}(x,\phi,\chi)+\sqrt{\frac{j+1/2}{j+3/2}}\,
\hat e_j^{m,q}\,  \mathcal N^{m-1,q+1}_{j+1}(x,\phi,\chi)\,,
\end{array}\end{equation*}
which gives the continuity of the operator $e^{-i\phi}\,e^{i\chi} (1+X)$ on both $\Phi_{H,F}$. Combining both last equations  we obtain an interesting expression:
\begin{equation}\begin{array}{l}\label{104}
e^{-i \phi}\,(\cos \chi + i\,x\,\sin\chi)\, \mathcal N^{m,q}_j(x,\phi,\chi) \\[2ex]\hskip0.75cm
= \frac12 \left[  \sqrt{\frac{j+1/2}{j-1/2}}  (c_j^{m,q}\, \mathcal N^{m-1,q-1}_{j-1} (x,\phi,\chi)     +\hat c_j^{m,q}\, \mathcal N^{m-1,q+1}_{j-1} (x,\phi,\chi))   \right. \\[2ex]\hskip1cm
\ds +\, d_j^{m,q}\, \mathcal N^{m-1,q-1}_{j} (x,\phi,\chi) + \hat d_j^{m,q} \, \mathcal N^{m-1,q+1}_{j}  (x,\phi,\chi) 
\\[2ex]\hskip1.25cm
 + \left. \sqrt{\frac{j+1/2}{j+3/2}} (e_j^{m,q} \, \mathcal N^{m-1,q-1}_{j+1} (x,\phi,\chi) + \hat e_j^{m,q}\, \mathcal N^{m-1,q+1}_{j+1}    (x,\phi,\chi)   )     \right]\,,
\end{array}\end{equation}
which  shows that the operator $e^{-i\phi} (\cos \chi+iX\,\sin\chi)$ is continuous on $\Phi_{H,F}$ and, hence, its formal adjoint is extended to  the duals $\Phi_{H,F}^\times$.

%%%%%%%%%%%%%%%%%%%%%%%%%
%%%%%%%%%%%%%%%%%%%%%%%%%
\section*{Appendix C: Inequalities for Jacobi functions.}\label{inequalities}

In formula (20) of Ref.~\cite{HS} it is given an  upper-bound for Jacobi polynomials  
  that in our notation is
\begin{equation}\label{c1}
|\mathcal J^{m,q}_j(x)| \leq  \left[\frac{(j-m+1)(j+m+1)}{(j-q+1)(j+q+1)}  \right]^{1/4}\,, 
\end{equation}
 valid for $x\in [-1,1]$,  $m\geq 0,\; m\geq |q|$ and $j\geq  |m|,\,|q|$. Moreover since
 \[
 (j-m+1)(j+m+1)-(j-q+1)(j+q+1)=-m^2+q^2\leq 0
\]
 we see that 
\begin{equation}\label{c2}
 |\mathcal J^{m,q}_j(x)|\leq 1\,, \qquad \forall m\geq 0,\, m\geq |q|,\, j\geq |m|, |q|\,.
\end{equation}
 
  In the following  let us consider only those Jacobi functions $\mathcal J^{m,q}_j(x)$ such that $(j,m,q)$ verify  that $m\geq q\geq 0$  and $j\geq m,q$. Then using that $\mathcal J^{m,q}_j(x)=\mathcal J^{q,m}_j(x)$ as we have proved in \eqref{a9} we see that inequality   \eqref{c2}  is valid for all $m,q\geq 0$. Morever,  according to 
  $\mathcal J^{m,q}_j(x)=(-1)^{m+q}\,\mathcal J^{-m,-q}_j(x)$ \eqref{a9} the inequality \eqref{c2}  is also valid for $(m,q)$ both negative numbers. 
  Since the inequality     \eqref{c2} is valid for those $\mathcal J^{m,-q}_j(x)$ with $m\geq q$ applying the same reasoning using the equalities \eqref{a9} we arrive to  proved the the inequality \eqref{c2} are valid
  for all $(j,m,q)$ verifying the conditions  displayed in \eqref{7}, i.e. 
  $j \ge |m|, |q|;\; 2j\in\mathbb N;\; j-m,\,   j-q\in\mathbb N$.

  On the other hand the inequality \eqref{c1} is valid for the same values of $(j,m,q)$ as \eqref{c2}
  with some small changes and now it reads as follows
  \begin{equation}\label{c3}
|\mathcal J^{m,q}_j(x)| \leq  \left[\frac{(j-a+1)(j+a+1)}{(j-b+1)(j+b+1)}  \right]^{1/4}\,, 
\end{equation}
where
$a=\max (|m|,|q|)$ and $b=\min (|m|,|q|)$. 

  As we mention before in Appendix A and taking into account the validity conditions of the \eqref{c1} we see that we have enlarged the inequality \eqref{c1} (or formula (20) of Ref.~\cite{HS})
 for all $\alpha$ and $\beta$ such that   $n\geq |\alpha|,|\beta|$.

Finally for the sake of the requirements of this work we can write the following upper-bound for the $\mathcal J^{m,q}_j(x)$
from \eqref{c3}
\begin{equation}\label{c4}
 |\mathcal J^{m,q}_j(x)|\leq (j+|m|+1)\,(j+|q|+1) \,, 
 \qquad \forall j \ge |m|, |q|;\; 2j\in\mathbb N;\; j-m,\,   j-q\in\mathbb N\,.
\end{equation}\end{widetext}

%%%%%%%%%%%%%%%%%%%%%%%%%%%%%%%%%%%%%%%%%%%%
%%%%%%%%%%%%%%%%%%%%%%%%%%%%%%%%%%%%%%%%%%%%

\end{document}